\documentclass[preprint]{elsarticle}
\usepackage[utf8]{inputenc}
\usepackage{graphicx}
\usepackage{color}
\usepackage{bm}
\usepackage{amsmath,amssymb}
\usepackage{natbib}

\newcommand{\basilisk}{{\usefont{T1}{pzc}{m}{n}Basilisk} }
\newcommand{\tr}{{\mbox{tr}}}

\begin{document}
\begin{frontmatter}

\title{An adaptive solver for viscoelastic incompressible two-phase problems applied to the study of the splashing of slightly viscoelastic droplets}
\author[1]{J.M López-Herrera}
\author[2]{S. Popinet}
\author[3]{A. A. Castrej\'on-Pita}
\address[1]{Departamento Ing. Aerospacial y Mecánica de Fluidos, Universidad de Sevilla, España}
\address[2]{Institut Jean Le Rond $\partial$'Alembert, Sorbonne Universit\'e, Centre National de la Recherche Scientifique, F-75005 Paris, France.}
\address[3]{Department of Engineering Science, University of Oxford, Oxford OX1 3PN, United Kingdom}

\begin{abstract}

We propose an adaptive numerical solver for the study of viscoelastic 2D two-phase flows using the volume-of-fluid method. The scheme uses the robust log conformation tensor technique of Fattal \& Kupferman \cite{Fattal2004,Fattal2005} combined with the time-split scheme proposed by Hao \& Pan \cite{Hao2007}. The use of this time-split scheme has been proven to increase the stability of the numerical computation of two-phase flows. We show that the adaptive computational technique can be used to simulate viscoelastic flows efficiently.  The solver is coded using the open-source libraries provided by the \basilisk \cite{Basilisk} platform. In particular, the method is implemented for Oldroyd-B type viscoelastic fluids and related models (FENE-P and FENE-CR). The numerical scheme is then used to study the splashing of weakly viscoelastic drops. The solvers and tests of this work are freely available on the \basilisk \cite{Basilisk} web site \cite{lopez}.
\end{abstract}
\end{frontmatter}

\section{Introduction}
Using numerical solutions for complex rheologies is nowadays a common predictive tool, since the efficiency of the numerical schemes improves continuously and the computational cost decreases. Typically, three main schemes have been used in computational fluid dynamics: Finite Differences (FD), Finite Volume (FV) and finite elements (FE). The presence of interfaces poses additional difficulties. Typical approaches to free surface simulations are the Marker and Cell (MAC), the Volume of Fluid (VoF) and the Level Set (LS) methods. The MAC method has been the reference method for numerous works since the pioneering work of Tomé et al (1996)\cite{Tome1996}.  Their original implementation of the MAC scheme is implemented within the framework of the FD method with the advection term approximated using the VONOS scheme\cite{Varonos1998}. The original implementation, conceived for simulating Oldroyd-B fluids, has been adapted to solve viscoelastic fluid of finite extensity as FENE-CR fluids \cite{Paulo2014}, using the log conformation kernel\cite{Martins2015} or the square root kernel \cite{PalharesJunior2016}.  Other numerical methods, such as the Smoothed-Particle-Hydrodynamics (SPH) method, can also be found in the literature on computational rheology\cite{Zainali2013,Xu2012}.

{\color{black}The Finite-Element method applied to viscoelastic flows goes back to the pioneering work of \cite{Nickell1974, Viriyayuthakorn1980, Guenette1995}. Successful implementations of viscoelastic fluids using FE have recently been conducted \cite{Kane2009, Knechtges2014} and is the basis of commercial codes as Polyflow\textregistered. The FE implementation of the log conformation schemes done by Hulsen et al. \cite{Hulsen2005} follows right after the original scheme is published. the FE implementation of the log conform performed by Hao \& Pan \citep{Hao2007} is particularly relevant for the present work since we use the time-split scheme proposed in that work.}

Most of the numerical simulations loose convergence and destabilize when the relaxation parameter, or its dimensionless counterpart, the Weissenberg number, is increased above a threshold value. This behaviour, known as the \textsl{High-Weissenberg number problem} (HWNP), has been a severe hindrance for computational rheology. Fortunately, a major relief of the HWNP problem has been provided by Fattal \& Kupferman \cite{Fattal2004,Fattal2005}. These authors proposed formulating the equations in terms of the logarithm of the conformation tensor. Interestingly, this log-conformation (kernel) formulation  guarantees the positive definiteness of the conformation tensor during the entire simulation. The success of this kernel method has been immediate, and is substituting, in practice, the classic approach in computational rheology. The log conformation kernel has been implemented within the FD method \cite{Fattal2004,Fattal2005}, the FE method \cite{Hulsen2005, Hao2007}  and the FV method \cite{Afonso2009}. In the same spirit \citep{Balci2011} proposed  using the square root of the conformation tensor to preserve the positive definiteness. Although less extended than the log conformation kernel, the square root conformation kernel has been used recently to analyze the lid cavity problem \cite{Dalal2016, PalharesJunior2016}. Although other conformation kernels are possible \cite{Afonso2012, Martins2015}, these seem to be the most accurate.

The FV is, at present, the method of reference in CFD (included commercial codes). Several reasons support its popularity. Remarkably, the method is intrinsically conservative, and the simulation of two-phase flows is straightforward since it does not require any special treatment.  Among the authors, contributing to the development of the FV scheme applied to viscoelastic flows, we can outline Alves and co-workers \cite{Alves2001, Alves2003}. These authors proposed a scheme consisting of a modification of the pressure-velocity coupling SIMPLEC algorithm combined with a new flux limiter for the advection term, CUBISTA, better suited for rheological fluids. Recently it has been shown that a streamfunction–log-conformation methodology\cite{Comminal2015,Comminal2016} can provide stable numerical simulations of flows with very high Weissenberg numbers. Figueiredo et al \cite{Figueiredo2016} have shown that the log-conformation formulation can be used together with the Continuum Surface Force method (CSF) to simulate accurately highly viscoelastic, surface tension dependent, two phase flows. Some implementations are constructed, profiting from existing CFD toolboxes, such as OpenFOAM\copyright \cite{Favero2010, Habla2011, Habla2014, Pimenta2017}. It is worth mentioning the efforts of some authors who put their codes at the disposal of the scientific community. This collaborative spirit allows a continuous improvement of the codes, as those done by Pimenta \& Alves \cite{Pimenta2017}.

Among the most common rheological models, we find the  Oldroyd-B\cite{Oldroyd}, Giesekus\cite{Giesekus}, FENE-type\cite{Bird80, Fene} or Phan-Thien-Tanner (PTT) \cite{Phan} models. Each of these models can better suit the particular solvent-polymer solution or melt employed in a particular problem. For example, either the Oldroyd-B or the FENE-type seems to fit properly the rheological behaviour of aqueous solutions of polyacrylamide (PAA) \cite{Purnode1998,Varagnolo2017}.
Both the FENE-P and FENE-CR  models correct the more simple Oldroyd-B model by imposing a maximum stretch that cannot be exceeded (FENE stands for Finitely Extensible Nonlinear Elastic), with the difference between them being the statistical closure used for the restoring force; P denotes the Peterlin's closure \cite{Bird80} and CR follows from the closure proposed by Chilcott \& Rallison\cite{Fene}. However, numerical simulations seldom match quantitatively the experiments in all of the possible regimes. Note, for example, that numerical simulations, using the Oldroyd-B model, have been employed successfully to explain the origin of the  ``beads on string" structure appearing in the breakup of weakly viscoelastic droplets \cite{Bhat2010}, but conversely, overestimate, largely, the damping factor in slightly vibrating pendant droplets \cite{Torres2017}.

We construct the viscoelastic solver using the free toolbox \basilisk developed by S. Popinet \cite{Basilisk}. Among the different solvers available in \basilisk we can find a library which deals with incompressible fluid problems with a second order in a space time-splitting projection method. Extra forces in the momentum equation can be easily included in the solver in a staggered way to avoid parasitic currents, and facilitate the balance of forces in steady equilibrium situations. The advection term in the momentum equation is computed using the Bell-Colella-Glaz (BCG) second order upwind method\cite{Bell1989}. The VoF method is used for two-phase flows with the advection of the interface performed using the conservative scheme of \cite{Weymouth2010}.  Surface tension forces are added using the Brackbill's CSF procedure \cite{Brackbill1992} in a balanced manner \cite{Popinet2009}. \basilisk also offers tools to easily perform an on-the-fly adaptation of the grid depending on the particularities of the flow studied. {\color{black} Adaptation has been used for viscoelastic fluids problems together with FE schemes \cite{Saramito2014,Jaensson2015}. Saramito \cite{Saramito2014} uses an anisotropic auto-adaptive mesh library to search efficient unstructured meshes capable to provide accurate stationary solutions to the lid cavity problem. In the method of Jaensson et al. \cite{Jaensson2015}, the grid moves with the fluid.   The mesh tends to become highly distorted and, in consequence, inaccurate. Jaensson er al. tackles the distortion by performing periodically a \emph{framing} and a remeshing as the computation proceed.}

On this platform we have implemented the classic viscoelastic approach in which the advancing equation is written in terms of the stress tensor. We have also implemented kernel conformation approaches, either the log conformation kernel of Fattal \& Kupferman, or the square root kernel of Balci et al. In all cases a time-split scheme is used with a calculation of the advection term with the BCG upwind scheme. For the log conformation kernel approach we go further with the time splitting by adopting the scheme of Hao \& Pan (2007). The constitutive model of reference in this work is Oldroyd-B, although for the kernel conformation approaches we have implemented also the FENE-P and the FENE-CR constitutive models for illustrative propose.

With these implementations we intend to (i) put at the disposal of the scientific community, a validated, ready-to-use, open-source solver using either the log conformation or the square root methodologies that can deal with multi-phase flows and fluids of complex rheology; (ii) gain insight on the advantages/drawbacks of the log conformation compared to the square root kernel in the case of two-phase flows; {\color{black} (iii) gain insight on the use of the adaption of grid in the resolution of viscoelastic two-phase problems} and (iv) report the results on the simulation of the spreading of a weakly viscoelastic fluid after its impact on a flat surface that can be either solid or a liquid layer or bath.

The impact of liquid droplets onto solid surfaces is present in many applications. Most of them search for a control of the coating of the solid by the fluid by managing the dynamics of the impacted droplets.  Many investigators have dedicated their efforts to this area of study when the fluid is Newtonian. A thorough review of the state-of-art research on this issue can be found in \cite{Josserand2016}. The addition of very small amounts of polymers to a solvent fluid enables a new degree of freedom for this control. In particular, it has been shown that very dilute polymeric solutions inhibit the rebound of droplets over hydrophobic surfaces \cite{Bertola2013}. In that article the impact dynamics of a droplet of water are compared with that doped with 200 ppm of Polyethylene Oxide (PEO). The spreading stage looks very similar for both Newtonian and viscoelastic fluids. The spreading is dominated by inertia, with negligible viscoelastic forces. Therefore, both droplets reach the same maximum width at the same time. However, the recoiling stage is much slower in the case of the doped droplet. Initially the slowdown of the receding contact line was attributed to the viscoelastic bulk phenomena in the vicinity of the contact line, but direct visualization has shown that the curbing is an interfacial phenomena between the substrate and the drop.  The contact line slows down because the polymer molecules are stretched perpendicularly to the contact line as the drop edge sweeps the substrate \cite{Bertola2013}. {\color{black} Recently Izbassarov \& Muradoglu (2016) \cite{Izbassarov2016} and Wang et al. (2017) \cite{Wang2017} has afforded numerically the study of the spreading and receding of impacting viscoelastic droplets. The authors of  \cite{Izbassarov2016} use a sharp interface scheme (front tracking) and set the contact angle \emph{ad hoc} at each computational step with the Kistler correlation. The numerical work described in \cite{Wang2017} is accomplished using the viscoelastic Giesekus model together with the diffuse-interface  Cahn-Hilliard model in which the interfaces are considered as thin transition regions where the interfacial forces are smoothly distributed. The objective of this work was to study the dynamic of the contact line (more precisely the dynamic of the contact angle) when viscoelasticity is present.}
Recently, \cite{Vega2017} have studied experimentally the dynamics of the splashing of slightly non-Newtonian drops onto a smooth surface. These authors pay special attention to the change on the map of the splashing modes (prompt splash, no splash or corona splash) due to the added polymer. They report that visco-elasticity hinders the development of prompt splashing.

The literature on the splashing on liquid baths is not as vast, as in the case of Newtonian fluids. \cite{Pregent2009} studied the splashing of viscoelastic droplets onto either Newtonian or viscoelastic baths. This issue is important, for example, for the formation of capsules or gelled beads.

This manuscript is organized as follows. In section \ref{ecuaciones} the governing equations of an isothermal and incompressible viscoelastic fluid are described. The log and the square root kernels are briefly derived. Details on the numerical schemes are given in section \ref{numerical}. Validation tests of the implemented numerical schemes are performed in section \ref{test}. In section \ref{splashing} we focus on the problem of the splash of weakly viscoelastic droplets.

\section{Governing equations}
\label{ecuaciones}
The equations governing the problem is the set formed by the mass
conservation equation,
\begin{equation}
\nabla \cdot \mathbf{u} = 0,
\end{equation}

and the momentum conservation,
\begin{equation}
\rho(\partial_t \mathbf{u} + \mathbf{u} \cdot \nabla \mathbf{u}) = - \nabla p + \nabla \cdot \bm{\tau} + \gamma \kappa  \mathbf{n} \delta_s +  \rho \mathbf{g}
\end{equation}
which relates inertia changes to, respectively,  the gradient of  pressures,
fluid internal stresses acting against deformation, surface tension forces
and, eventually, gravitational forces. We denote the density, velocity,
pressure and surface tension, curvature(which is normal to the interface $\mathbf{n}$) by $\rho$,
$\mathbf{u}$, $p$,  $\gamma$ and $\kappa$. $\delta_s$ stands
for the Dirac delta being one at the interface and zero elsewhere. The fluid
internal stresses are usually split into the solvent part, $\bm{\tau_s}$, and
the polymeric (viscoelastic) contribution $\bm{\tau_p}$,
\begin{equation}
\bm{\tau = \tau_s + \tau_p},
\end{equation}
while the solvent stress part depends on the deformation tensor as expressed for a usual Newtonian fluid,
\[\bm{\tau_s} = 2 \mu_s \mathbf{D} = \mu_s (\nabla \mathbf{u} + \nabla
\mathbf{u}^T) \, ,
\]
And the polymeric stress, $\bm{\tau_p}$, takes into account memory effects of the
polymers. Several constitutive rheological models are available
in the literature with their polymeric stresses $\bm{\tau_p}$, which are typically functions $\mathbf{f_S}(\cdot)$ of the
conformation tensor $\mathbf{A}$,
\[
\bm{\tau_p} = \frac{\mu_p \mathbf{f_S}(\mathbf{A})}{\lambda} \label{fr}
\]
where $\lambda$ is the relaxtation parameter of the fluid and $\mu_p$ the polymeric viscosity.  The conformation tensor $\mathbf{A}$  can be regarded as an internal state variable measuring the molecular deformation of the polymer chains \cite{Carreau1991}. The conformation tensor $\mathbf{A}$ is assumed to be always symmetric and positive definite, obeying the equation
\begin{equation}
\overset{\triangledown}{\mathbf{A}} = -\frac{\mathbf{f_R}(\mathbf{A})}{\lambda} \label{conf}
\end{equation}
where $\mathbf{f_R}(\mathbf{A})$ is the relaxation function which is different for each particular constitutive model. $\overset{\triangledown}{}$  denotes the operator
\textsl{upper-convected derivative} given by
\begin{equation}
 \overset{\triangledown}{\mathbf{A}} = \partial_t \mathbf{A} +   \nabla \cdot (\mathbf{u} \mathbf{A}) - \mathbf{A} \cdot \nabla \mathbf{u} - \nabla \mathbf{u}^{T} \cdot \mathbf{A}
\end{equation}
with $\nabla \mathbf{u}|_{ij}  = \partial_i u_j$.  $T$ denotes the
``transverse'' tensor. In table \ref{fsyr} the expressions of the strain and relaxation functions for some constitutive models are gathered.

\begin{table}
  \centering
  \begin{tabular}{lcccc}
                    & Oldroyd B      & FENE-P & FENE-CR & linear PTT\\ \hline
  $\mathbf{f_R(A)}$ & $\mathbf{A-I}$ & $\frac{\mathbf{A}}{1-\tr(\mathbf{A})/L^2}-\mathbf{I} $ & $\frac{\mathbf{A-I}}{1-\tr(\mathbf{A})/L^2}$ & $(1 +\varepsilon\, \tr(\mathbf{A-I}))(\mathbf{A-I})$\\
  $\mathbf{f_S(A)}$ & $\mathbf{A-I}$ & $\frac{\mathbf{A}}{1-\tr(\mathbf{A})/L^2}-\mathbf{I}$  & $\frac{\mathbf{A-I}}{1-\tr(\mathbf{A})/L^2}$& $\mathbf{A-I}$\\
  \hline
\end{tabular}
  \caption{Strain and relaxation functions, $\mathbf{f_S(A)}$ and $\mathbf{f_R(A)}$, for some constitutive models\cite{Comminal2015}. $\tr(\mathbf{A})$ stands for the trace of the tensor $\mathbf{A}$.   }\label{fsyr}
\end{table}

Classically, in the case of the Oldroyd-B model, it is usual to skip the use of $\mathbf{A}$ by combining Eqs (\ref{fr}) and (\ref{conf}). Then the constitutive equations in terms of the viscoelastic stress tensor, $\bm{\tau_p}$ writes,
\begin{equation}
 \lambda \bm{\tau_p} + \overset{\triangledown}{\bm{\tau_p}} = 2 \mu_p \mathbf{D} \label{classic}
\end{equation}

\subsection{The kernel conformation transformation}

The numerical resolution of viscoelastic problems often fails to converge when
the relaxation parameter, $\lambda$, is larger than relatively low values.
This instability has been termed in the literature the
\textsl{High-Weissenberg number problem} (HWNP), and it has been a major
obstacle in computational rheology. Fattal \& Kupferman
\cite{Fattal2004,Fattal2005} identified that the instability was caused by a
defective modelling of the exponential growths of the stresses. When the instability manifests itself the conformation tensor no longer maintains its property of being definite positive.
To tackle the HWNP  matrix kernel-transformations of the original conformation tensor have been proposed to enforce at every instant the positive-definite character of the tensor. Two main kernels transformations have been proposed: the log-conformation of  Fattal \& Kupferman
\cite{Fattal2004,Fattal2005} and the square-root-conformation of Balci et al. \cite{Balci2011}.

\subsubsection{Log conformation}
In this kernel, due to Fattal \& Kupferman, rather than advancing
the conformation tensor, they suggest to advance in time its logarithm,
$\bm{\Psi} = \log \mathbf{A}$. Note that, since $\mathbf{A}$ is symmetric and
positive-definite, and it is always diagonalizable, then,
\begin{equation}
\mathbf{A} = \mathbf{R} \,  \Lambda \,  \mathbf{R}^T \quad \mbox{and} \quad \Psi = \log \mathbf{A} = \mathbf{R} \, \log \Lambda \,  \mathbf{R}^T
\end{equation}
where $\Lambda$ is the diagonal matrix formed with the eigenvalues and $\mathbf{R}$ is the tensor formed by arranging the eigenvectors.

The diagonalization can also be used to decompose the velocity gradient as
\begin{equation}
(\nabla \mathbf{u})^T = \bm{\Omega} + \mathbf{B} + \mathbf{N} \mathbf{A}^{-1} \label{dec}
\end{equation}
where $\bm{\Omega}$ and $\mathbf{N}$ are antisymmetric and $\mathbf{B}$ is symmetric, traceless and commutes with $\mathbf{A}$. Using the above decomposition the equation for $\bm{\Psi}$ is,
\begin{equation}
\partial_t \bm{\Psi} + \mathbf{u} \cdot \nabla \bm{\Psi} - 2 \mathbf{B} -
(\bm{\Omega} \bm{\Psi} -  \bm{\Omega} \bm{\Psi}) = -\frac{e^{-\Psi}}{\lambda} \mathbf{f_R}(e^{\Psi}) \label{logeq}
\end{equation}
with homogeneous Neumann boundary conditions for $\bm{\Psi}$ by default.

In 2D the decomposition (\ref{dec}) is straightforward. In the case of zero
polymeric stresses $\bm{\tau}_p=0$, the elements
of the decomposition are, $\bm{\Omega}=0$ and $\mathbf{B}
=\frac{1}{2}[(\nabla \mathbf{u})^T+(\nabla \mathbf{u})]$. Otherwise, given
the diagonalized conformation tensor
\begin{equation}
\mathbf{A} = \mathbf{R}
\left(
\begin{array}{cc}
\Lambda_1 &  0 \\
 0 & \Lambda_2  \\
\end{array}
\right)\mathbf{R}^T
\end{equation}
the velocity gradient is written as
\begin{equation}
\left(
\begin{array}{cc}
m_{11} &  m_{12} \\
 m_{21} & m_{22}  \\
\end{array}
\right)
=\mathbf{R}^T (\nabla \mathbf{u})^T  \mathbf{R}
\end{equation}
and the elements of the decomposition as
\begin{equation}
\begin{split}
&\mathbf{B} = \mathbf{R} \left(
\begin{array}{cc}
m_{11} &  0 \\
 0 & m_{22}  \\
\end{array}
\right) \mathbf{R}^T, \,
\bm{\Omega} = \mathbf{R} \left(
\begin{array}{cc}
0 &  \omega \\
 -\omega & 0  \\
\end{array}
\right) \mathbf{R}^T \mbox{and} \\ &\mathbf{N} = \mathbf{R} \left(
\begin{array}{cc}
 0 &  n \\
 -n & 0  \\
\end{array}
\right) \mathbf{R}^T \quad \mbox{with} \quad \omega= \frac{\Lambda_2 m_{12} + \Lambda_1 m_{21}}{\Lambda_2 - \Lambda_1} \quad \mbox{and} \quad n = \frac{m_{12} + m_{21}}{\Lambda_2^{-1} - \Lambda_1^{-1}}.
\end{split}
\end{equation}
Expressions for the 3D case have been derived in \cite{Habla2014}. {\color{black} The square root kernel methodology of Balci et al.\cite{Balci2011} as well as details of its numerical time integration are briefly described in \ref{square_root} while for the classic approach, details are described in \ref{classical_mumerical}.}

\section{Numerical scheme}
\label{numerical}
We have built the numerical scheme using as a basis the open-source code
\basilisk\cite{Basilisk}. \basilisk provides both
ready-to-use Finite Volume (FV) solvers for fluid dynamics problems
(shallow-water, compressible, incompressible, multi-phase...), and an ensemble of useful
\textsl{c-language} libraries in order that users can tailor, with a
moderate effort, their own specific code.

The incompressible \basilisk solver uses a second order in space time-splitting projection method. The interface is tracked with a color
variable, $c(\mathbf{x}, t)$, which represents the volume fraction. $c$ is
convected with the fluid,
\begin{equation}
\partial_t c + \mathbf{u} \cdot \nabla c = 0 \, .
\end{equation}

\begin{figure}
  \centering
  \includegraphics[width=14cm]{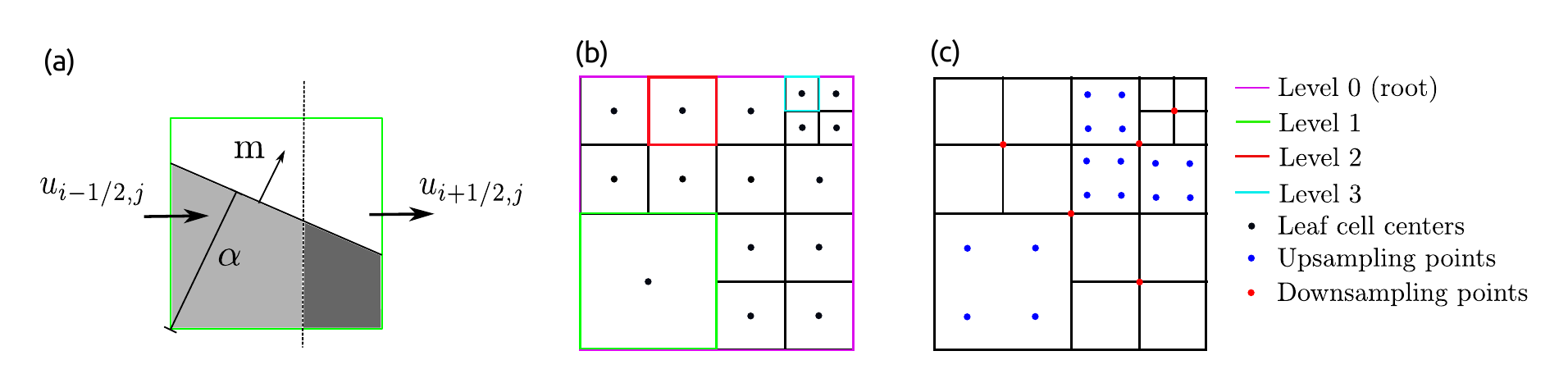}\\
  \caption{(a) Advection scheme of the volume fraction. (b) Quadtree structure (c) Location of the upsampling and downsampling points for the adaption.}\label{fig1}
\end{figure}

{\color{black} The above volume fraction equation is solved by successively advecting (sweeping) $c$ along each of the spatial directions, $x$ and $y$ (or $r$ in cylindrical coordinates), using a one-dimensional scheme.  As it is depicted in figure \ref{fig1}.a, the one-dimensional flux along the sweeping direction is computed from the local linear reconstructed equation, $\mathbf{m} \cdot \mathbf{x} = \alpha$, and the face velocities. This one-dimensional net flux must be corrected in case that the one-dimensional velocity field were not divergence-free, i.e, $u_{i-1/2,j} \neq u_{i+1/2,j}$ in figure \ref{fig1}.a. We use the dilation correction proposed in \cite{Weymouth2010} which it has been proved to be simple, robust  completely volume conservative (if the velocity field is divergence-free). The direction of the first of the one-dimensional sweeps is swapped between $x$ and $y$ in each computational step to avoid preferred direction of advection.}

The surface tension stresses are  added to the momentum 
equation with the CSF method\cite{Brackbill1992} in a balanced manner which avoid parasitic
currents\cite{Popinet2009}. The curvature of the interface is computed accurately using the height function approach. {\color{black} 
In this method the curvature is calculated using the height functions in horizontal or vertical direction, $x=h_x(y)$ and $y=h_y(x)$, being the curvature $\kappa$ (say in an almost horizontal interface) given by,
\begin{equation*}
\kappa = \frac{h_y''}{\sqrt{1+h_y'^2}} \, .
\end{equation*}

If the interface is almost vertical, $\kappa$ can be calculated similarly with $x=h_x(y)$ instead of $y=h_y(x)$. The method allows to obtain second-order accurate estimates of the curvature. The limits of resolution of the method appear when the size of the cell $\Delta$ is such that $\kappa \Delta \simeq 1$. No special treatment is required in this method for interfacial cells (cells in which the interface is located) next of boundaries and walls. A more detailed description of the method as well as a revision of the state of the art in the numerical calculation of surface tension stresses is available in \cite{Popinet2009, Popinet2018}.
}

The time stepping of the Navier-Stokes equations is as
follows
\begin{enumerate}[Step 1:]
\item The volume fraction is advanced in time using a conservative, non-diffusive geometric VoF,
\begin{equation}
\frac{c^{n+1/2}- c^{n-1/2}}{\Delta t} = \mathbf{u}^{n} \cdot \nabla c^{n}
\end{equation}
\item Polymeric stresses are advanced to mid-step $n+1/2$, $\bm{\tau_p}^{n+1/2}$.
\item Fluid properties are updated,
\begin{equation}
\theta^{n+1/2} = \theta_1 c^{n+1/2} +\theta_2 (1-c^{n+1/2})
\end{equation}
where $\theta$ stands for any property of the fluid; i.e, $\rho$, $\mu_s$, $\mu_p$ and $\lambda$ with subscripts 1 and 2  representing the bulk property at each phase.
\item An estimation of the velocity,  $\mathbf{u}^{*}$, is calculated by solving
\begin{equation}
\begin{split}
&\frac{\mathbf{u}^{*} - \mathbf{u}^{n}}{\Delta t} + \mathbf{u}^{n+1/2} \cdot \nabla \mathbf{u}^{n+1/2}
 = \\ &\frac{1}{\rho^{n+1/2}} \left(- \nabla p^n + \nabla \cdot (2 \mu^{n+1/2} \mathbf{D}^*) + \nabla \cdot \bm{\tau_p}^{n+1/2} + \gamma \kappa^{n+1/2} \nabla c^{n+1/2} \right) \label{momentum}
 \end{split}
 \end{equation}
where the advection term is calculated using the Bell-Colella-Glaz second order upwind scheme.
\item The velocity field is projected,
\begin{equation}
\nabla \cdot \left( \frac{\Delta t}{\rho^{n+1/2}} \nabla p^{n+1} \right) = \nabla \cdot \mathbf{u}^{*}
 \end{equation}
and updated,
\begin{equation}
\mathbf{u}^{n+1} = \mathbf{u}^{*} - \nabla p^{n+1} \, \Delta t/\rho^{n+1/2}
 \end{equation}
\end{enumerate}
The time step, $\Delta t$, is determined from two constraints; the stable explicit advection, which implies that the Courant-Friedrich-Levy (CFL) number is below 0.5, and the absence of fake capillary waves which obliges it to have $\Delta t \le (\rho h^3)/(\pi \gamma)$.

\subsection{Time integration of the polymeric stresses using the log conform kernel}
Although we use the log-conformation approach of Fattal \& Kupferman we still
use as a main variable the polymeric stress tensor, $\bm{\tau_p}$, as this has
been proposed by Figueiredo et al.\cite{Figueiredo2016}. Also, in the present scheme we apply the time-split procedure of \cite{Hao2007} in which Eq. (\ref{logeq}) is decomposed as
\begin{gather}
\partial_t \bm{\Psi}  + \mathbf{u} \cdot \nabla \bm{\Psi} = 0 \label{advection}  \\
\partial_t \bm{\Psi}  - 2 \mathbf{B} - ( \bm{\Omega} \bm{\Psi} - \bm{\Psi} \bm{\Omega}) = 0 \label{upper} \\
\partial_t \bm{\Psi} = \frac{e ^{-\bm{\Psi}}\mathbf{f_R}(e ^{\bm{\Psi}})}{\lambda}  \,. \label{const}
\end{gather}
Given the polymeric stresses at time $n - 1/2$, $\bm{\tau_p}^{n - 1/2}$, and the velocity field at instant $n$, $\mathbf{u}^{n}$ a generic time step proceeds as follows:
\begin{enumerate}[Step 1:]
\item The corresponding conformation tensor at instant $n - 1/2$ is calculated from the relationship,
\begin{equation*}
 \bm{\tau_p}^{n - 1/2} = \frac{\lambda}{\mu_p} \mathbf{f_S} (\mathbf{A}^{n - 1/2})
\end{equation*}
We assume that the stress function,  $\mathbf{f_S} (\mathbf{A})$, and the relaxation function, $\mathbf{f_S} (\mathbf{A})$, are linear functions
\begin{equation*}
\mathbf{f_{S,R}} (\mathbf{A}) = \eta_{S,R} (\nu_{S,R} \mathbf{A} - \mathbf{I})
\end{equation*}
For example, for the FENE-P constitutive model the parameters would be  $\eta_{S} = \eta_{R} = 1$ and $\nu_{S} = \nu_{R} =1/(1-Tr(\mathbf{A}^{n - 1/2})/L^2) $.
\item The conformation tensor is diagonalised, $\mathbf{A} = \mathbf{R} \,  \Lambda \, \mathbf{R}^T $, to obtain its eigenvalues and eigenvectors matrix,  $\Lambda^{n - 1/2}$ and $\mathbf{R}^{n - 1/2}$.
\item The log of the conformation tensor is calculated,
\begin{equation*}
\bm{\Psi}^{n - 1/2} = \left.\mathbf{R} \,  \log (\Lambda) \, \mathbf{R}^T \right|^{n - 1/2}
\end{equation*}
\item The gradient velocity is decomposed accordingly to  Eq. (\ref{dec}) to obtain $\mathbf{B}^n$ and $\bm{\Omega}^n$. Note that for the decomposition we use the eigenvalues and eigenvectors values at instant $(n-1/2)$.
\item The log-conformation tensor is advected using the BCG scheme, 
\begin{gather*}
\bm{\Psi}^*  =  \bm{\Psi}^{n-1/2} - {\Delta t} \nabla \cdot (\mathbf{u}^n  \bm{\Psi}^n)
\end{gather*}
\item Eq. (\ref{upper}) can be integrated explicitly,
\begin{gather*}
\bm{\Psi}^{**} = \bm{\Psi}^{*} + \Delta t ( 2 \mathbf{B}^n +  \bm{\Omega}^n \bm{\Psi}^{n-1/2} - \bm{\Psi}^{n-1/2} \bm{\Omega}^n )
\end{gather*}
or, implicitly,
\begin{gather*}
\bm{\Psi}^{**} = \bm{\Psi}^{*} + \Delta t ( 2 \mathbf{B}^n +  \bm{\Omega}^n \bm{\Psi}^{**} - \bm{\Psi}^{**} \bm{\Omega}^n)
\end{gather*}
Note that an implicit integration could easily be accomplished given that the resulting equations are linear, and the unknowns at a given point are uncoupled from the unknowns at neighboring points. It would consist in solving $N$ times, once per grid point, a linear system of 3 unknowns (in cartesian 2D; $\Psi_{xx}$, $\Psi_{yy}$ and $\Psi_{xy}$). Our numerical tests on this issue suggest that nothing is gained with the implicit integration.
\item The constitutive model Eq. (\ref{const}) is written in terms of the conformation tensor,
\begin{equation*}
\partial_t \mathbf{A} = -\frac{\mathbf{f_R} (\mathbf{A}) }{\lambda}
\end{equation*}
and later integrated analytically.

\begin{enumerate}
\item Prior to the analytical integration, the log of the conformation tensor is diagonalised,
\begin{equation*}
\bm{\Psi}^{**} = \left.\mathbf{R} \,  \log (\Lambda) \, \mathbf{R}^T \right|^{**}
\end{equation*}
to obtain $\Lambda^{**}$, $\mathbf{R}^{**}$ and the conformation tensor, $\mathbf{A}^{**} = \left.\mathbf{R} \, \Lambda \, \mathbf{R}^T \right|^{**}$.
\item Then,   $\mathbf{A}^{n+1/2}$ is calculated with
\begin{equation*}
\mathbf{A}^{n+1/2} = \mathbf{A}^{**} \, e^{-\eta_R \nu_R \Delta t/ \lambda} + (1- e^{-\eta_R \nu_R\Delta t/ \lambda}) \frac{\mathbf{I}}{\nu_R}
\end{equation*}
\end{enumerate}
\item Finally,
\begin{equation*}
\bm{\tau_p}^{n + 1/2} = \frac{\mu_p}{\lambda} \mathbf{f_R}(\mathbf{A}^{n + 1/2})= \frac{\mu_p}{\lambda} \eta_R (\nu_R \mathbf{A}^{n + 1/2}  - \mathbf{I})
\end{equation*}
\end{enumerate}

\subsection {Spatial discretization and the adaption algorithm}
{\color{black}The open code \basilisk discretizes the computational domain using a structured grid of square finite volumes (termed hereafter \textsl{cells}) that can be either uniform or non-uniform. If a non-uniform grid is preferred, the discretization is arranged hierarchically in a quadtree structure\cite{Popinet2003} (see figure \ref{fig1}.b). In this type of structure, the size of a cell, $h$, is characterized by its \textsl{level}, $\ell$, at which is located. Hence, the size of the cells at that level $h \varpropto 2^{-\ell}$.  A prototypical cell of level $\ell$ can be \textsl{parent} of 4 \textsl{children} cells (at the level $\ell +1$).  The \textsl{root} cell is that corresponding to $\ell = 0$ from which the rest of the cells at a higher level hang down. A \textsl{leaf} cell is a cell without any child. In the example shown in figure \ref{fig1}.b, the grid would be formed by 16 leaf cells being four of them of level $\ell=3$, one of level $\ell=1$ and the rest of level $\ell=2$.  All the main variables, including the components of the polymeric stress tensor $\bm{\tau_p}$, are defined at the cell center. However, the stresses of the right side of the momentum equation (\ref{momentum}) are computed at the cell faces to avoid  any spurious current that could result from the imbalance between pressure and elastic stresses.

This tree-type grid structure allows the performance of a fast and efficient do-loop across the grid nodes. Besides, adding a few constraints in the growth of the tree branches, as for example that the maximum jump of level between neighbouring leaf cells is one, the grid can be refined and coarsened dynamically (\textsl{adapted}) as the simulation proceeds at an affordable computational cost. The adaptation is based in a multi-resolution analysis of selected scalar fields. Consider a control scalar field  discretized at grid level $\ell$, $f_\ell$. This scalar field can be coarsened to the lower level by means of a downsampling operation denominated \emph{restriction},
\begin{equation}
f_{\ell-1} = restriction(f_\ell)\,.	
\end{equation}
This coarser field distribution, $f_{\ell-1}$, can be upsampled (or \emph{prolongated}) to the original level, 
\begin{equation}
g_{\ell} = prolongation(f_{\ell-1})\, ,	
\end{equation}
and compared to the original distribution to provide an estimation of the error, $\xi_\ell = ||f_\ell-g_\ell||$. Given a particular cell $i$ of level $\ell$ in which the error is $\xi^i_\ell$, then that cell will be,
\begin{itemize}
	\item Refined if $\xi^i_\ell > \zeta$,
	\item Coarsened if $\xi^i_\ell < 2\zeta/3$,
	\item Remain unchanged otherwise.
\end{itemize} 
where $\zeta$ is the error threshold set. The prolongation procedure is second-order accurate and involves additional upsampling points in cells contiguous to the finer ones (see figure \ref{fig1}.c). A more detailed explanation of the adaption algorithm can be found in \cite{vanHooft2018}. Observe that to fill the new refined and coarsened cells with proper values for each variable can be done with inter/extrapolations that could differ of the prolongation and restriction operators used to decide adaption regions. In our experience, it is better to use as control adaption variables the velocity components and the volume of fraction. The values of  $\bm{\tau_p}$ in the new refined cells are computed with a bilinear interpolation while in the coarser ones they are calculated by averaging. }

\begin{figure}
	\centering
	\includegraphics[width=\linewidth]{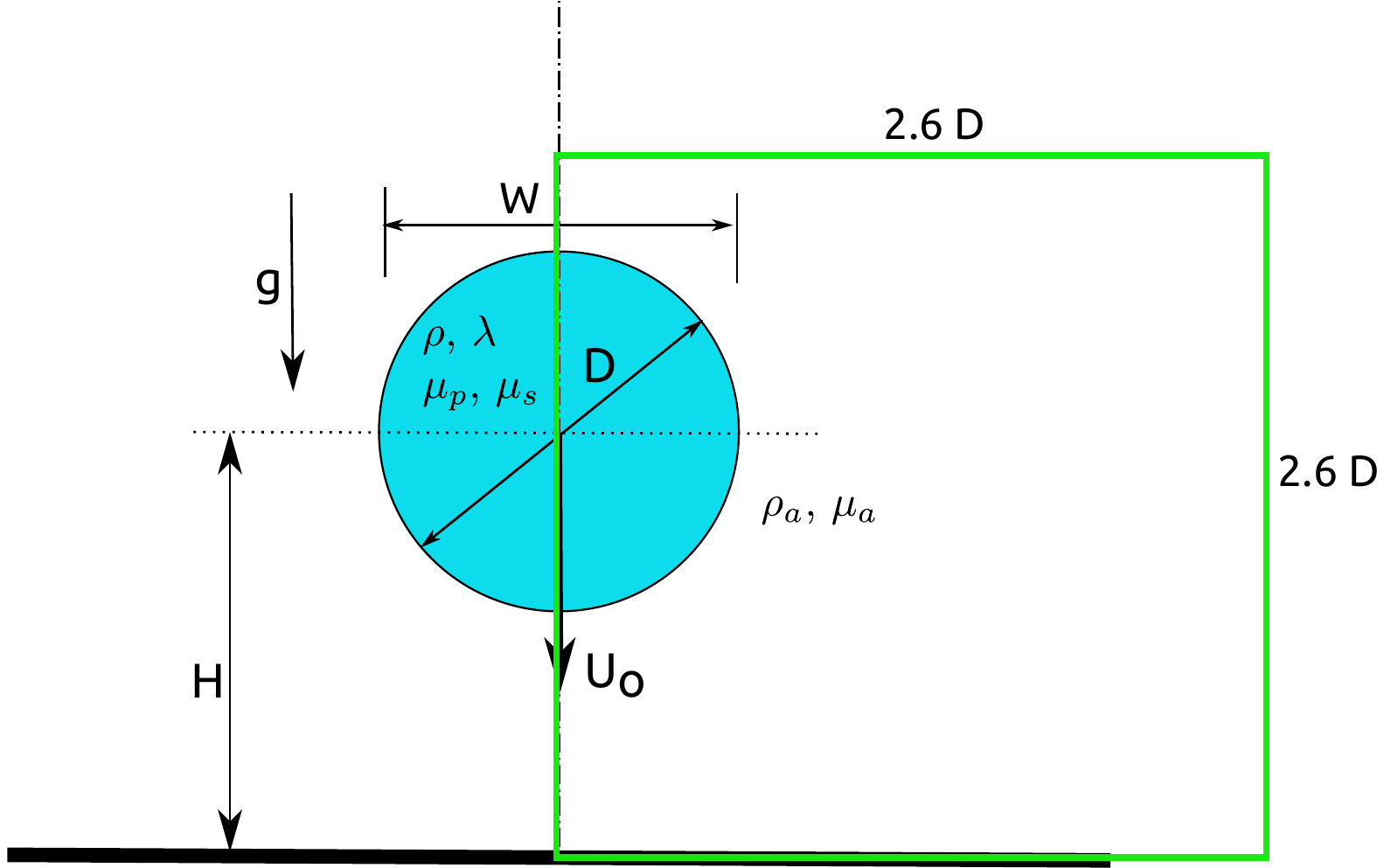}\\
	\caption{Sketch of the problem. The green square denotes the computational domain.}\label{fall}
\end{figure}

{\color{black} As a result of the hyperbolic nature of the equations for $\mathbf{A}$, boundary conditions ought to be only considered at inflows\cite{VanderZanden1988} where we impose by default homogeneous Neumann boundary conditions for tensors, $\bm{\Psi}$, $\mathbf{b}$ and $\bm{\tau_p}$. However, since in our numerical scheme all the viscoelastic stress components are defined at the centers of the cells, some care must be taken to suitably model the presence of walls and symmetries in the momentum equation (\ref{momentum}). Note that the viscoelastic force density applies in our scheme at cell faces and requires to set values at ghost cells since the force density is calculated using central differences. The values at the ghost cells follow the expressions derived in section 3 and 4 of \cite{Tome2007}. In the case of a rigid wall of orientation $\mathbf{n}$ , the normal component, $\bm{\tau}_{\bm{p},nn}$, would be zero. Note that $\bm{\tau}_{\bm{p},nn} = 0$ is only valid for certain constitutive models. For the axisymmetric case, the boundary condition, $\bm{\tau}_{\bm{p},\theta \theta} = 0$ must be added at the wall. Also, on the axis of symmetry the conditions
\[
\partial_r \bm{\tau}_{\bm{p},\theta \theta} = \partial_r \bm{\tau}_{\bm{p},r r} = \partial_r \bm{\tau}_{\bm{p},zz} = 0, \quad \text{and} \quad \bm{\tau}_{\bm{p},rz} = 0 \, .
\]
must be imposed.  
}

\section{Test}
{\color{black} We have performed various test of the numerical schemes presented in this work to verify aspect as the time integration or the correct treatment of the interaction of the viscoelastic fluid with walls and interfaces. Those tests unrelated specifically to the splashing problem are gathered in  \ref{test}.}
	
\subsection{Splashing of a viscoelastic droplet}
\label{drop_fig}
This test case is intended to validate the code for axisymmetric two-phase flows in the absence of surface tension. Additionally, some insight in adaptation is gained. The study deals with the time evolution of a viscoelastic Oldroyd-B droplet of density $\rho$, relaxation parameter, $\lambda$, solvent and polymeric viscosity, $\mu_s$ and $\mu_p$, and diameter $D$ launched from a height $H$ at a velocity $U_o$ as sketched in figure \ref{fall}. The surrounding atmosphere is assumed to be dynamically negligible, i.e. $\rho_a \rightarrow 0$ and $\mu_a \rightarrow 0$. The scaling of the equations of motion will be carried out with the liquid density, $\rho$, the droplet diameter $D$ and the fall velocity $U_o$ to give a Froude number, $Fr = gD/U^2_o$, a dimensionless height $h = H/D$, a Reynolds number $Re = \rho D U_o/(\mu_p +\mu_s)$, a Deborah number $De = \lambda U_o/D$, the ratio of solvent to total viscosity $\beta = \mu_s/(\mu_s + \mu_p)$, and the ratio of the outer to inner density and viscosity,  $\rho_r = \rho_a/\rho$ and $\mu_r = \mu_a/\mu_o$, respectively.  This test case has been used by diverse authors with very different schemes \cite{Figueiredo2016, Xu2012}. As in the previous work of \cite{Figueiredo2016} the dimensionless parameters were fixed to: $Fr = 2.26$, $h = 2$, $De =1$, $Re = 5$ and $\beta = 0.1$. \cite{Figueiredo2016} do not report values for the outer medium; in the present work we set either $\mu_r$ and  $\rho_r$ to $10^{-3}$. In what follows the dimensional variables are denoted by an asterisk.

We have simulated these tests using the log kernel, the square root kernel and the classic methodology. The computational domain in the present simulations is also shown in Figure \ref{fall}. It consists of a square of dimensionless size $2.6 \times 2.6$. We use axisymmetric equations with the left boundary as the axis of symmetry.  The mesh in the simulations is adapted depending on the components of dimensionless velocity, $u_x$, $u_y$ and volume fraction, $f$. We have set two ensemble of threshold values; $\varepsilon^{th}_{f} = \varepsilon^{th}_{u_x} = \varepsilon^{th}_{u_y} = 10^{-3}$ (adaption A1) and $\varepsilon^{th}_{f} = 10^{-3}$ and  $\varepsilon^{th}_{u_x} = \varepsilon^{th}_{u_y} = 10^{-2}$ (adaption A2). The simulation performed by \cite{Figueiredo2016} were made with uniform meshes ranging from $\Delta r = \Delta r^*/D= \Delta z = \Delta z^*/D = 2.5 \, \times \, 10^{-2}$ up to  $\Delta r =  \Delta z = 1.25 \times 10^{-2}$. Since in \cite{Figueiredo2016} negligible difference between meshes are shown, For both adaptation strategies, A1 and A2, the cell widths are comprised between $\Delta r = \Delta z = 2.03 \,\times\, 10^{-2}$ and $\Delta r = \Delta z = 8.12 \,\times\, 10^{-2}$.  The maximum timestep has been fixed in all simulations to $\Delta t = U_o \Delta t^*/D = 10^{-3}$.

\begin{figure}
  \centering
  \includegraphics[width=\linewidth]{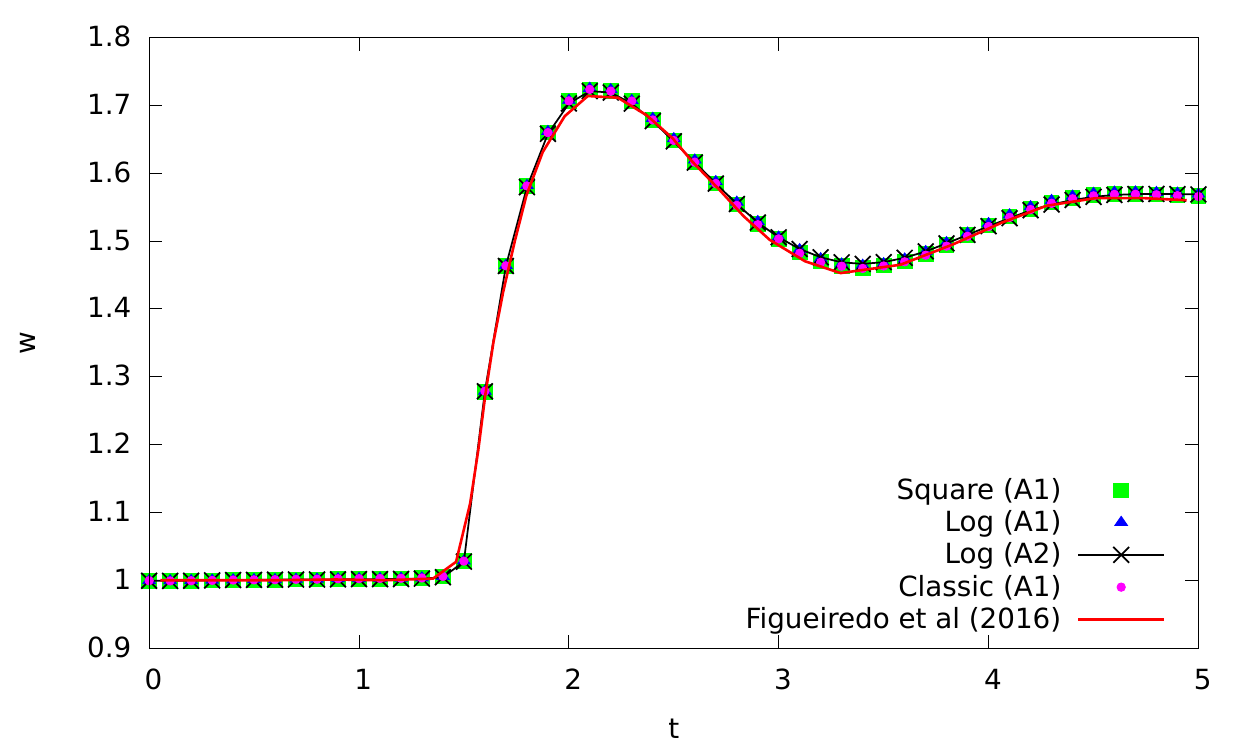}\\
  \caption{Time evolution of the dimensionless width of the droplet, $w$. Numerical simulation using adaptation strategy A1: (i) Eq. (\ref{logeq}) (ii) Eq. (\ref{eqbalci}) and (iii) Eq. (\ref{classic}). The numerical simulation of \cite{Figueiredo2016} is also shown (continuous red line). The results with the adaptation strategy A2 and the log kernel methodology are also shown.}\label{fall2}
\end{figure}

Figure \ref{fall2} shows the dimensionless width of the droplet, $w = W/D$ versus the dimensionless time $t = t^* U_o/D$. We compare our results with the different methodologies against those found in \cite{Figueiredo2016} with the adaptation strategy A1. All three methodologies give very consistent results and are in very good agreement with the results of \cite{Figueiredo2016}.

\begin{figure}
  \centering
  \includegraphics[width=\linewidth]{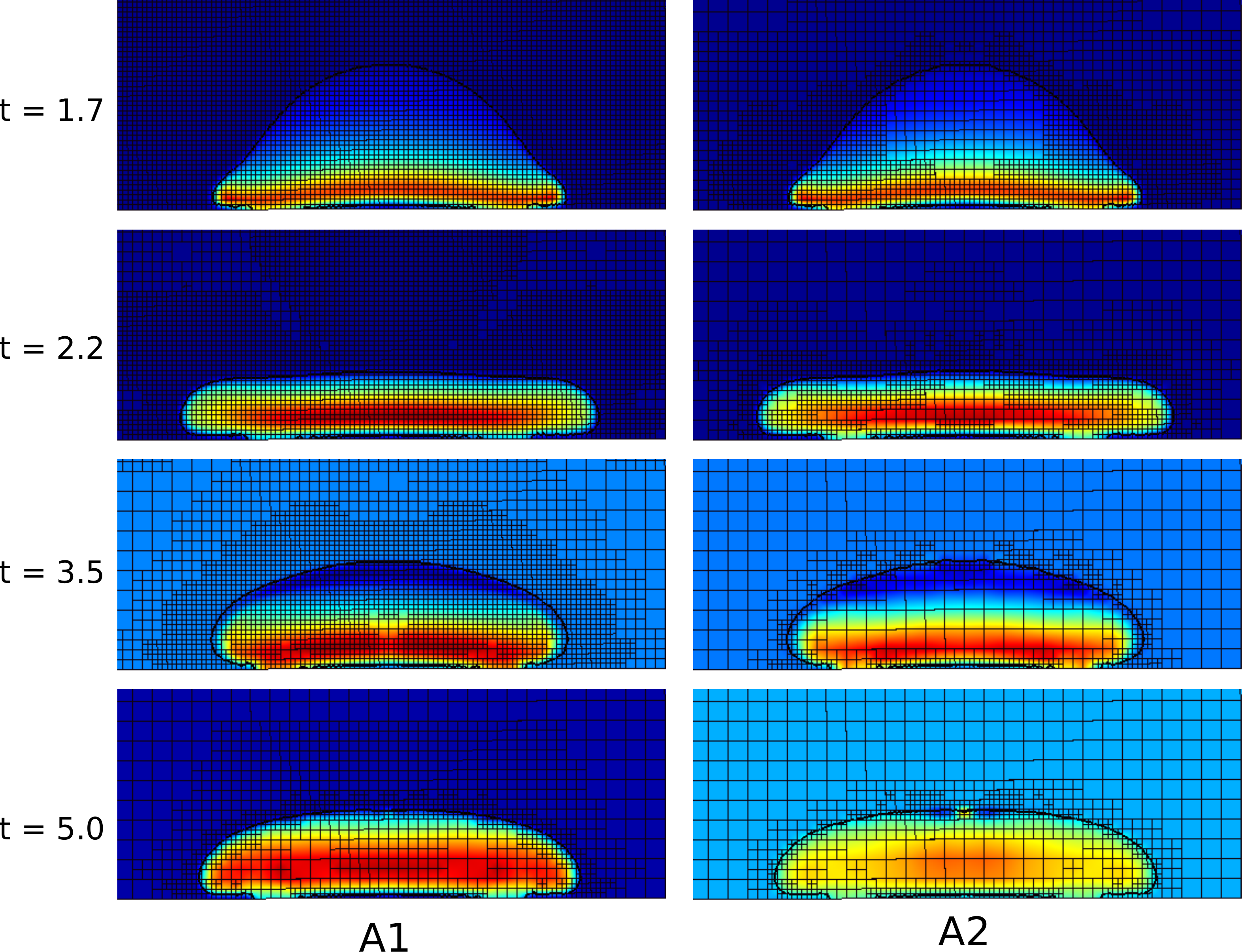}\\
  \caption{Snapshots at instant $t$ =1.7, 2.2, 3.5 and 5 with adaptation A1 and A2. Each snapshot shows the mesh, the interface position and the spatial distribution of the component of the stress tensor, $\bm{\tau}_{\bm{p},\theta \theta}$}\label{fall3}
\end{figure}

\section{Splash of weakly viscoelastic drops}
\label{splashing}
In this section we investigate the splash of a viscoelastic drop onto flat substrates. The substrate can be either solid or a viscoelastic liquid film/bath. The properties of the viscoelastic fluid used in the simulations correspond to those of mixtures of pure distilled water with small quantities (around 0.01 wt\%) of polymeric solutions of polyacrylamide (PAA), as in the experiments of Vega \& Castrejon-Pita \cite{Vega2017}. Table \ref{properties} shows the dependence of the viscoelastic properties, $\mu_p$ and $\lambda$, on the solution concentration. The solvent properties are those of distilled water, $\mu_s = 10^{-3}$ Pa s and  $\rho = 998$ Kg/$\mbox{m}^3$. The surface tension is unaffected by the polymeric additives, and is therefore equal to $\sigma = 0.072$ N/$\mbox{m}$. As in subsection \ref{drop_fig} we use as scaling magnitudes the liquid density $\rho$, the droplet impact velocity $U_o$ and the droplet diameter $D$ (we set $D$ = 3.28 \, mm as in the experiments of \cite{Vega2017}). Therefore, a particular splashing is characterized by the following dimensionless quantities:

\begin{itemize}
\item A global Reynolds number, $Re = \rho D U_o/(\mu_s +\mu_p)$.
\item A Weber number, $We = \rho D U^2_o/\sigma$. Sometimes, in the literature, instead of $We$ the splashing parameter, $K= We \sqrt{Re}$ is used.
\item A Deborah number $De = \lambda U_o/D$ and a ratio of solvent to total viscosity $\beta = \mu_s/(\mu_s + \mu_p)$.
\item The ambient to solvent properties ratios, $\mu_r = \mu_a/\mu_s$ and  $\rho_r = \rho_a/\rho$.
\item {\color{black}The dimensionless height at which the droplet is released, $H/D$.}
\item Finally, if the substrate is a liquid film of width $L^*$, its relative depth $L = L^*/D$.
\end{itemize}

Note that in the above list of parameters the Froude number, $Fr= U_o/\sqrt{g D}$, is absent because it is irrelevant in the splashing phenomena ($Fr \gg 1$) despite the fact that gravity plays a crucial role for accelerating the droplet up to the impact velocity $U_o$. Also other parameters that can be relevant, such as the contact angle or the aspect ratio of the droplet before the impact, are not explored.

The numerical simulation is performed using axisymmetric equations in a square domain similar to the one depicted in figure \ref{fall}. Adaptation is performed at each timestep according to the velocity field and the interface position. The simulations have been carried out with different degrees of grid refinement. Most simulations have been carried out with a grid as fine as 5461 cells per diameter in the adapted region, while far away of that area the grid is coarsened to an equivalent of 21 cells per droplet diameter.  Occasionally, for the largest falling velocities, the finest grid reached an equivalent of 10922 cells per diameter. In a few selected cases, the simulations have been performed on parallel machines.

\subsection{Solid substrate}

When the substrate is a solid, the simulation can be started shortly after the impact of the droplet. As shown by \cite{Philippi2016}, the computed dynamics of the spreading of the droplet, using a slightly truncated landed sphere as initial geometry, is entirely similar to the one obtained while releasing the droplet in the air. While an air dimple can be created when releasing the droplet in air, it does not affect the dynamics of the spreading lamella \cite{Philippi2016}.
We have selected to initiate the simulations with the the center of the sphere located at a dimensionless distance $H/D = (1-5 \,\times\, 10^{-5})$ above the substrate {\color{black} being the downward dimensionless velocity of the viscoelastic fluid uniform and equal to $u_z = -1$. The rest of the variables are set to zero.}

To explore the influence of the viscoelasticity on the overall dynamics, we focus on the splash of a 1000 ppm solution droplet at an impact velocity $U_o = 4.09$ m/s, that corresponds to $We=760$. For this concentration the other parameters take the following values; $Re = 576.33$, $De=174.51$ and $\beta=0.043$. We impose the ratios values, $\mu_r = 0.018$ and $\rho_r=0.001$. For comparison purposes we also simulate the Newtonian case of a pure solvent (0 ppm).

\begin{figure}
  \centering
  \includegraphics[width=\linewidth]{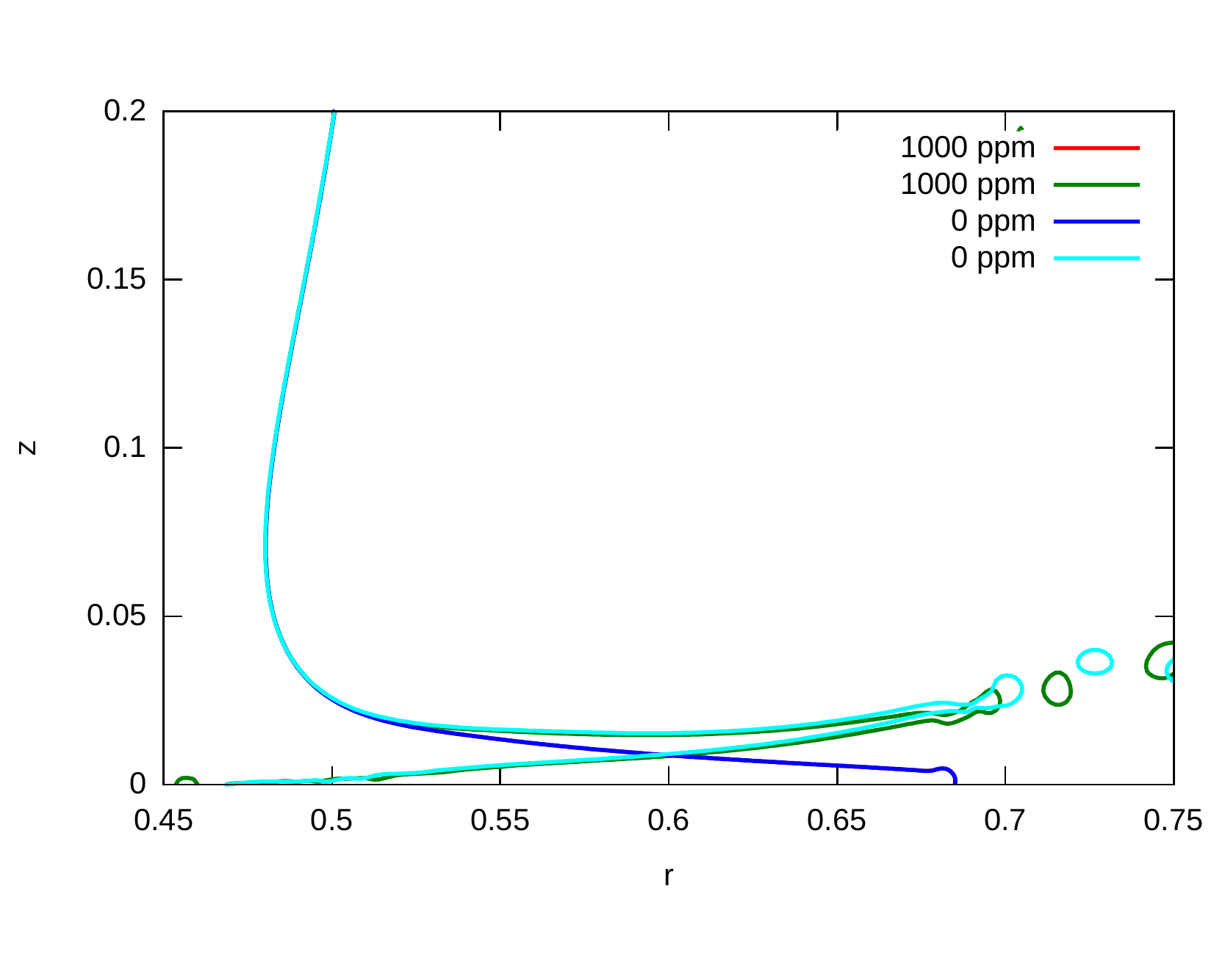}\\
  \caption{\color{black}Detail of the shape and position of the splashed sheet and lamella at $t$  = 0.18 for solutions 0 ppm (Newtonian) and 1000 ppm.  In the case of the levitated sheet, the boundary condition at the wall for the volume concentration, $c$, is $c=1$. For the sliding lamella the normal derivative of $c$ at the wall is nullified, $\partial_n c = 0$.} \label{fig12}
\end{figure}

\begin{figure}
  \centering
  \includegraphics[width=\linewidth]{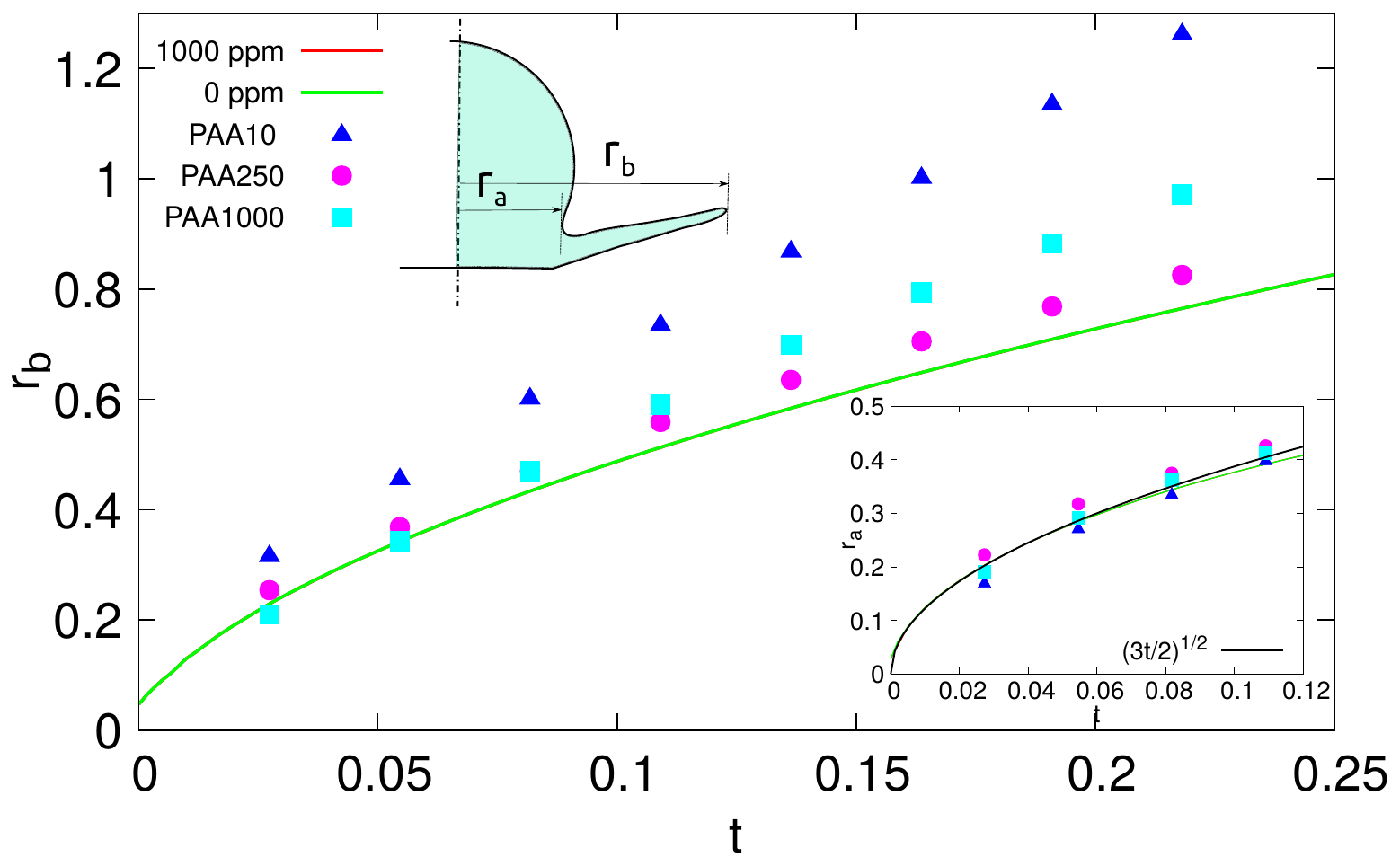}\\
  \caption{Time evolution of the spreading of the droplet after splashing. The dimensionless lamella tip radius, $r_b$, is plotted versus the dimensionless time for 1000 ppm and 0 ppm concentrations at $We=760$ (continuous lines). {\color{black} Experimental points of Fig. 9 in Vega \& Castrejon-Pita \cite{Vega2017} are also shown.} The radial position of the turning point, $r_a$, is shown in the inner figure and compares the numerical simulations with Wagner's analytical solution, $r_a = \sqrt{3t/2}$ and the experimental results of \cite{Vega2017}.} \label{fig13}
\end{figure}

Figures \ref{fig12} and \ref{fig13} show details of the droplet splashing for both 0 and 1000 ppm in concentration. In order to investigate the effect of the wall-fluid interaction we have set a different boundary condition of the volume fraction , $c$, at the wall. The effect of wall-fluid interaction is shown in Figure \ref{fig12} where we plot the shape of the lamella at instant $t=0.18$. The sliding lamella (red and black lines) is obtained with the default boundary condition of zero normal derivative, $\partial_n c = 0$. This condition corresponds to a contact angle of $\pi/2$. The levitating sheet is obtained by imposing the Dirichlet condition, $c = 1$ (green and cyan lines). Interestingly, the elastic effects are negligible for these very low polymer concentrations as it can be observed in Figures \ref{fig12} and \ref{fig13}. The mechanism of the splashing is unaffected by the viscoelastic character of the fluid, at least for the very small concentrations cases.{\color{black} Interestingly, the experiments performed by Jung et al.\cite{Jung2013} on the splashing of droplets of solutions of polystyrene in diethyl phthalate over a highly wettable solid  exhibits the same irrelevance of the polymer concentration in the dynamic of the splashing}. Note that both the lamella tip radius, $r_b$, and the radial position of the turning point, $r_a$, are not affected by the viscoelastic stresses {\color{black}in the numerical simulations}. Furthermore, the calculated position $r_a$ fits well with the analytical Wagner solution obtained using potential theory \cite{Philippi2016}, similarly to the experiments of Vega \& Castrejon-Pita \cite{Vega2017}. This matching suggests that, in the bulk of the fluid, either viscous and viscoelastic stresses are unimportant during the first stages of impingement. Viscous and viscoelastic effects are confined to the wall boundary layer and along the contact line. Our numerical results suggest that the viscoelasticity could alter the contact line equilibrium that, in turn, affects the dynamics of the lamella. {\color{black} This numerical result agrees well with the experimental results of \cite{Jung2013} where the wettability of the fluids (with and without polymers) is so high that the contact angle is not longer a relevant parameter and the spreading of the droplet is unaffected by the presence of polymers. This is not the general case since the substrate will play a relevant (non simple) role in the spreading and receding stages as the dynamic contact angle will vary in the process\cite{Wang2017,Izbassarov2016}.}

\begin{table}
  \centering
  \begin{tabular}{ccc}
              \hline
     Solution concentration (ppm) & $\mu_p$ (Pa s) & $\lambda$ (s) \\ \hline
     100 & 2.22 $\,\times\,10^{-3}$ & 0.0086 \\
     250 & 5.55 $\,\times\,10^{-3}$ & 0.0196 \\
     1000 & 2.22 $\,\times\,10^{-2}$ & 0.14 \\             \hline
            \end{tabular}
  \caption{Polymeric viscosity, $\mu_p$, and relaxation time, $\lambda$, in S.I units for different diluted PAA-water solutions \cite{Vega2017}.}\label{properties}
\end{table}

\subsection{Liquid substrate}

When the substrate is liquid, the droplet is released at a height equal to $H/D= 1.05$ setting the dimensionless velocity $u_z = - 1$ as an initial condition to all the fluid in the droplet. {\color{black}The rest of variable are initially zero}. The thickness of the film layer has been set to $L=0.3$, which seems to be enough to simulate splashing in a deep pool, since simulations done with thicker film layers than $L=0.3$ do not show any difference in the mechanism and shape of the splashing. We have simulated splashing with $We$ ranging from 50 up to 760 that correspond to falling velocities of 1.05 m/s up to 4.09 m/s for a droplet diameter of 3.28 mm, respectively.

\begin{figure}
  \centering
  \includegraphics[width=\linewidth]{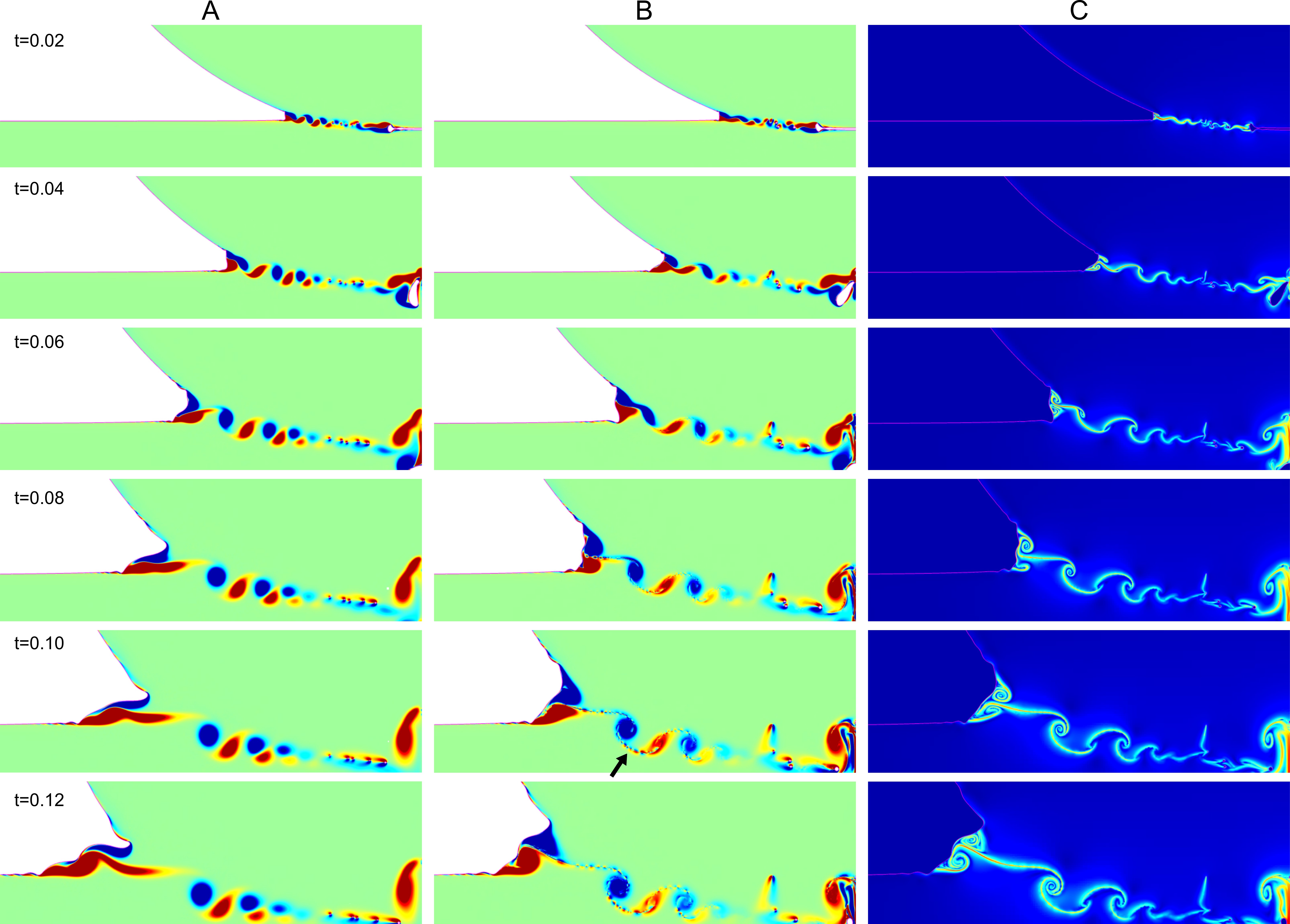}\\
  \caption{ Snapshots of the splashing of a droplet with $We=50$.  From top to  bottom at instants  $t=0.02$ to $t= 0.12$ in steps of $\Delta t = 0.02$ . Column A shows the vorticity distribution, $\omega$, for 0 ppm. Column B shows the vorticity distribution for 1000 ppm. Column C shows the distribution of the $\ell2$ norm $||\Psi||_2$ (Eq. (\ref{Psi2})) for 1000 ppm.} \label{fig14}
\end{figure}

Figure \ref{fig14} shows the first stages of the splashing for $We=50$ for the pure Newtonian case of 0 ppm and the slightly viscoelastic fluid case of 1000 ppm. In the figure we plot the vorticity distribution $\omega$ given by
\begin{equation}
\omega = \frac{\partial u_r}{\partial z} - \frac{\partial u_z}{\partial r} \, .\label{vort}
\end{equation}
Figure \ref{fig14} also shows the $\ell2$-norm of the conformation tensor $\Psi$,
\begin{equation}
||\Psi||_2 = \sqrt{\Psi_{rr}^2  + \Psi_{zz}^2 +  \Psi_{\theta \theta}^2  + 2 \Psi_{rz}^2 } \, .\label{Psi2}
\end{equation}
$||\Psi||_2$ is used to visualize where the viscoelastic stresses are more intense. As can be seen in the figure \ref{fig14}, and the supplementary material, as the drop squeezes the film, the junction front between the drop and film advances and thickens rapidly. In its advance the front flaps, as a consequence of the vortex shedding, creating a Von Kármán-type vortex street, as was already pointed out by Thoraval et al. \cite{Thoraval2012} and confirmed experimentally by \cite{Castrejon2012}. At the same time the gas entrapped in the dimple, formed between the droplet and film, rapidly retracts to form a bubble. At the first stages (snapshots $t=0.02$ and $t=0.04$) no apparent difference exists in the vorticity distribution between the 0 and the 1000 ppm mixtures. However, in subsequent stages it can be observed that the vortex pairs are more distant for the case of a viscoelastic drop (column B) compared to the Newtonian one (column A), since the viscoelastic stresses slightly drag out the shedding of vortices.

The evolution of these vortical structures is more interesting. In the case of a Newtonian fluid the vortical structures can only decay by viscous diffusion of the momentum. Since splashing characteristic times are short, and the Reynolds number is large ($Re = 3432$ for the Newtonian fluid of figure \ref{fig14}), the vorticity distribution within the bulk of the liquid is practically the same in snapshot $t=0.06$ and subsequent ones. In the case of the mixture of 1000 ppm, the picture is altered by the viscoelastic stresses. Generally speaking, the viscoelastic stresses disrupt this vortical structure as time goes by. Between the spots of positive-negative vorticity, which form the paired vortex, a trail of alternated micro-vortices appears (shown by the black arrow in the fifth snapshot of column B). Note that, in this case, the spots of vorticity rapidly loose their homogeneity decaying  in a turbulent-like mixing.

\begin{figure}
  \centering
  \includegraphics[width=\linewidth]{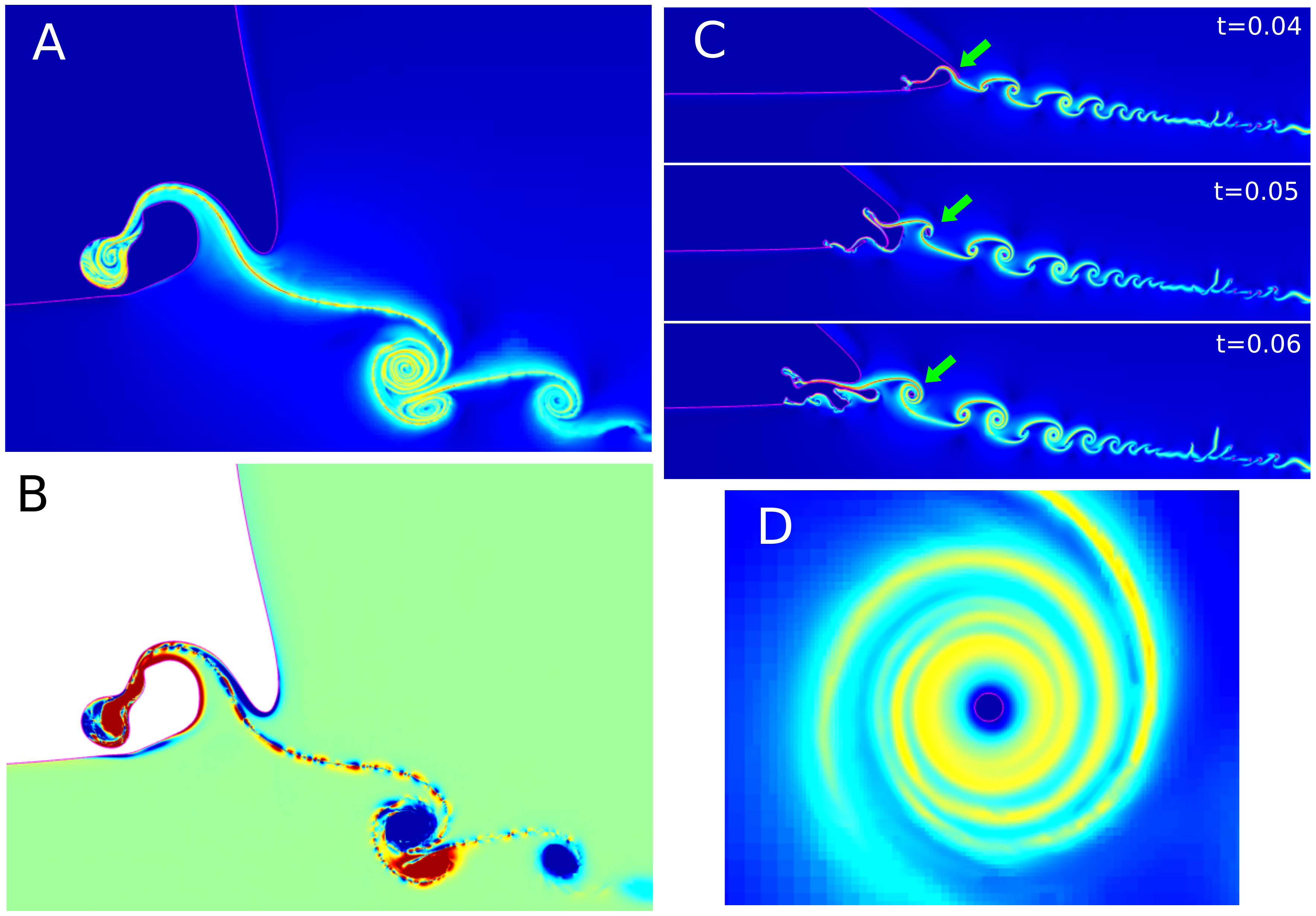}\\
  \caption{Figures A and B correspond to a 1000 ppm fluid falling with $We=300$. A and B show $\omega$ and $||\Psi||_2$ distributions  at instant $t = 0.2$, respectively. Figure C shows the process of entrapment of a ring bubble and the subsequent roll up of viscoelastic stresses for a 1000 ppm liquid with $We = 760$. Figure D shows the details of the structure of the roll up of viscoelastic stresses around ring bubbles. } \label{fig15}
\end{figure}

Viscoelastic stresses are concentrated on the fluid surface separating the fluid of the drop from the fluid of the pool, since it is there that larger deformation and strain occur during the splashing process. As can be observed in figure \ref{fig15}A, in the lamella, a central core sheet of viscoelastic stresses acts against its spreading and development. In some cases, particularly for violent high $We$ number splashes, the viscoelastic stresses tend to bend the incipient lamella, making the first stages of the splashing highly chaotic, as can been seen in sequence \ref{fig15}C. Interestingly, the vortices roll up the viscoelastic stresses giving some sort of toroidal spring that delays the advance of the lamella (see figures \ref{fig15}A,B and D). These structures are particularly intense when generated around ring bubbles. A sequence of the nucleation of a toroidal spring around a bubble is shown by the green arrow in figure \ref{fig15}C. In the first snapshot we can see how the flapping lamella entraps a bag of air by hitting the falling droplet. This bag of air, already has a bubble ring, and is rotating, straining the fluid, and rolling up this strained viscoelastic fluid (second snapshot). Finally, a toroidal spring-like structure is the result. Details of this structure are shown in figure \ref{fig15}D.

\section{Conclusions}

In this article we have shown how the time-splitting scheme proposed by Hao \& Pan \cite{Hao2007} can be used together with the classical log conformation tensor of Fattal \& Kupferman  \cite{Fattal2005}, or the square-root conformation of Balci et al. \cite{Balci2011}, to provide stable numerical simulations of two-phase viscoelastic flows. It is also shown that the time-splitting scheme simplifies the extension of the numerical scheme to different constitutive laws with a moderate effort. Many of the numerical results presented here have been obtained using adaptivity, which can be applied straightforwardly to viscoelastic simulations. The solvers, and most of the tests performed in the present study, are freely available on the \basilisk web page \cite{lopez}.

The numerical scheme has been used to investigate numerically the splashing of weakly viscoelastic droplets on to solid flat substrates and pools of the same fluid, taking as reference the experimental conditions of the work of Vega \& Castrejon-Pita \cite{Vega2017}.
We observe no difference in the splashing process onto hard substrate between pure solvent droplets and slightly viscoelastic droplets because the viscoelastic bulk effects are negligible for the polymer concentration used.  Therefore, we hypothesize that the differences observed by Vega \& Castrejon-Pita are due to alterations of the contact line equilibrium because of viscoelasticity and that, in turn, affects the advance of the lamella.

In contrast, the splashing of a slightly viscoelastic droplet onto a pool exhibits a phenomena that has not already been observed in Newtonian fluids. We have observed that the viscoelastic stresses alter the vortex shedding, reported by Thoraval et al. \cite{Thoraval2012}. Also, as the splashing proceeds a trail of alternated micro-vortices appears. The viscoelastic stresses are responsible for the disruption of these vortices. The shedding vortices strain the fluid with its rotation, and rolls up this strained viscoelastic fluid to form some sort of toroidal spring. These toroidal springs can nucleate around trapped bubble rings similar to those reported by \cite{Thoraval2012}.

\label{conclusion}

\section*{Acknowledgments}
This work has been supported by the Spanish Ministry of Economy Grant DPI2013-46485. J. L-H wishes to thank M.-J. Thoraval for his guidance in the analysis of droplet splashing. AACP was funded through an EPSRC-UK grant (EP/P024173/1) and a Royal Society University Research Fellowship.

\appendix

\section{Square-Root conformation}
\label{square_root}
\subsection{Equations}
Balci et al.\cite{Balci2011}  propose to formulate the constitutive differential
models in terms of the (unique) positive symmetric square root $\mathbf{b}(x, t)$ of the conformation tensor $\mathbf{A}(x, t)$,
\[
\mathbf{A} = \mathbf{b}\mathbf{b}^T
\]
that substituted in Eq. (\ref{conf}) the results in the following time advancing equation for $\mathbf{b}$,
\begin{equation}
\partial_t \mathbf{b} +   \nabla \cdot (\mathbf{u} \mathbf{b}) = \mathbf{b} \cdot \nabla \mathbf{u} + \mathbf{a} \mathbf{b} -\frac{\mathbf{b}^{-1} \, \mathbf{f_R}(\mathbf{b}\mathbf{b}^T)}{\lambda}
\label{eqbalci}
\end{equation}
where $\mathbf{a}$ is an antisymmetric tensor in which off-axis values result from the enforcement of the symmetric character of $\mathbf{b}$. In 2D this would be
\[
\mathbf{a} = \left(
\begin{array}{cc}
0 &  a_{12} \\
-a_{12} & 0  \\
\end{array}
\right) \quad \text{being} \quad a_{12} = \frac{b_{12}\partial_x u_x -b_{11} \partial_x u_y +
	b_{22} \partial_y u_x - b_{12} \partial_y u_y}{b_{11} + b_{22}}
\]
Balci et al. \cite{Balci2011} provide expressions for the 3D case.
\subsubsection{Numerical scheme}
As for the case of the log kernel the numerical scheme is a time splitting procedure of Eq. (\ref{eqbalci}). Therefore, a time step can be decomposed in the following substeps.
\begin{enumerate}[Step 1:]
	\item The square root tensor is advected explicitly with the BCG scheme,
	\begin{equation*}
	\mathbf{b}^{*} = \mathbf{b}^{n-1/2} + \Delta t \, \nabla \cdot (\mathbf{b}^{n} \mathbf{u}^n)
	\end{equation*}
	\item The rest of Eq. (\ref{eqbalci}) is linearized and solved implicitly. Assuming a linear relationship for the relaxation function, the system to be solved would be,
	\begin{equation*}
	\frac{\mathbf{b}^{n+1/2}}{\Delta t} -
	\mathbf{b}^{n+1/2} \nabla \mathbf{u}^{n} -
	\mathbf{a}^{n} \cdot \mathbf{b}^{n+1/2} + \frac{\eta_R \nu_R}{\lambda}
	\mathbf{b}^{n+1/2} = \frac{\mathbf{b}^{*}}{\Delta t} + \frac{\eta_R}{\lambda} (\mathbf{b}^{-1})^{n-1/2}
	\end{equation*}
	\item Finally the polymeric stress is computed from $\mathbf{b}^{n+1/2}$
	\begin{equation*}
	\mathbf{A}^{n+1/2} = \mathbf{b}^{n+1/2} \mathbf{b}^{T,n+1/2} \quad \text{and} \quad \bm{\tau_p}^{n + 1/2} = \frac{\mu_p}{\lambda} \mathbf{f_S} (\mathbf{A}^{n+1/2})
	\end{equation*}
\end{enumerate}

\section{Numerical scheme for the classic approach}
\label{classical_mumerical}
In this scheme we solve Eq. (\ref{classic}) by time splitting. The step procedure is as follows.
\begin{enumerate}[Step 1:]
	\item The stress components are advected explicitly with the BCG scheme,
	\begin{equation*}
	\bm{\tau_p}^{*} = \bm{\tau_p}^{n-1/2} + \Delta t \, \nabla \cdot (\bm{\tau_p}^{n} \mathbf{u}^n)
	\end{equation*}
	\item The upper convective derivative is solved implicitly,
	\begin{equation*}
	\left(1 + \frac{\lambda}{\Delta t} \right) \bm{\tau_p}^{n+1/2} -  (\nabla \mathbf{u}^T)^n \, \bm{\tau_p}^{n+1/2}
	+ \bm{\tau_p}^{n+1/2} \nabla \mathbf{u}^n  = 2 \mu_p D + \lambda \frac{\bm{\tau_p}^*}{\Delta t}
	\end{equation*}
\end{enumerate}

\section{Additional tests}
\label{test}
\begin{figure}
	\centering
	\includegraphics[width=0.75\linewidth]{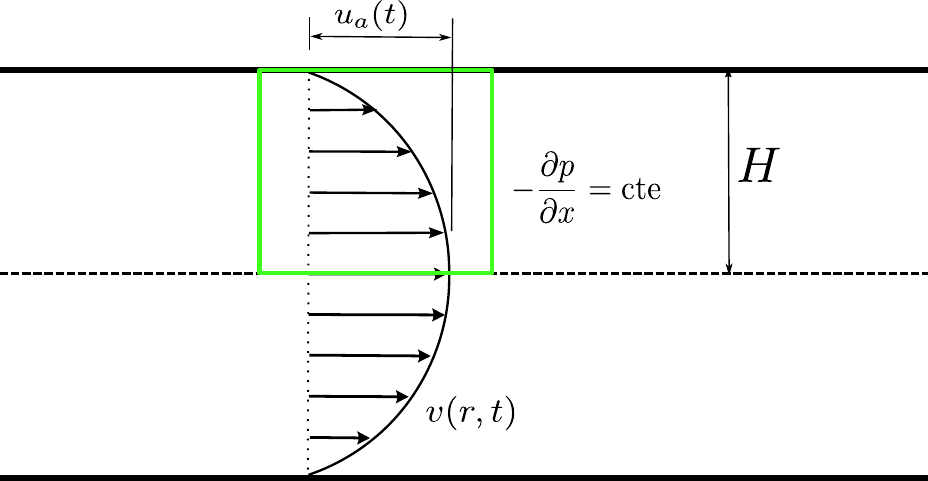}\\
	\caption{Sketch of the transient onset planar Poiseuille flow for a viscoelastic fluid. The simulation domain is depicted in green.}\label{fig2}
\end{figure}

\subsection{Transient planar Poiseuille flow for a viscoelastic fluid }
The problem is sketched on figure \ref{fig2}. A viscoelastic fluid of density $\rho$, solvent and polymeric viscosity, $\mu_s$ and $\mu_p$, and relaxation time $\lambda$, is trapped in the gap of width $2H$ formed by two parallel infinite plates. The fluid, initially at rest, is set in motion by the sudden application of a constant pressure gradient. The steady planar parabolic Poiseuille flow is reached after a transient period if the viscoelastic fluid is an Oldroyd-B or a FENE-CR. For a FENE-P the profile departs slightly from a strict parabola \cite{Oliveira2002}. The transient for a Newtonian fluid is characterized by an exponential increase of the axial velocity. However, in the case of a viscoelastic fluid its elastic nature gives a different behaviour, since an oscillation is superposed to the exponential increase. For an Oldroyd-B fluid, an analytical solution due to Waters \& King \cite{Waters1971} is available,
\begin{equation}
u(y,t) = 1.5(1-y^{2}) - 48 \sum_{k=1}^\infty \frac{\sin((1+t)n/2)}{n^3}
e^{\alpha_n t/2} G(t) \label{pois_anal}
\end{equation}
with $n=(2k-1)\pi$, $\alpha_n = 1 + \beta \,E \, n^2 /4$ and
\[
G(t) =\sinh(\theta_n t/2) + \frac{\gamma_n}{\theta_n} \cosh(\theta_n t/2)
\]
with
\[
\theta_n = \sqrt{\alpha_n^2 - E \, n^2} \quad \mbox{and} \quad
\gamma_n = 1 - \frac{2-\beta}{4} \,E \, n^2
\]
where $E$ is the elastic number given by $E = \lambda \mu_o/(\rho H^2)$ and $\beta$ is the ratio of the solvent to total viscosity, $\beta= \mu_s/\mu_o = \mu_s/(\mu_s +\mu_p)$. In the analytical expression (\ref{pois_anal}), the time $t$ and the position $y$ are dimensionless magnitudes. They result after the time is made dimensionless in $\lambda$, $t = t^*/\lambda$, the y-coordinate with $H$, $y = y^*/H$ and the velocity with the average steady velocity,
\[
u(y,t) = \frac{u^*(y^*,t^*)}{\bar{u}^*_\infty} \quad \text{being} \quad  \bar{u}^*_\infty = -\frac{\Delta p^*}{\Delta x^*} \frac{H^2}{3 \mu_o}.
\]
where the superscript $^*$  denotes the dimensional counterpart.

\begin{figure}
	\centering
	\includegraphics[width=0.8\linewidth]{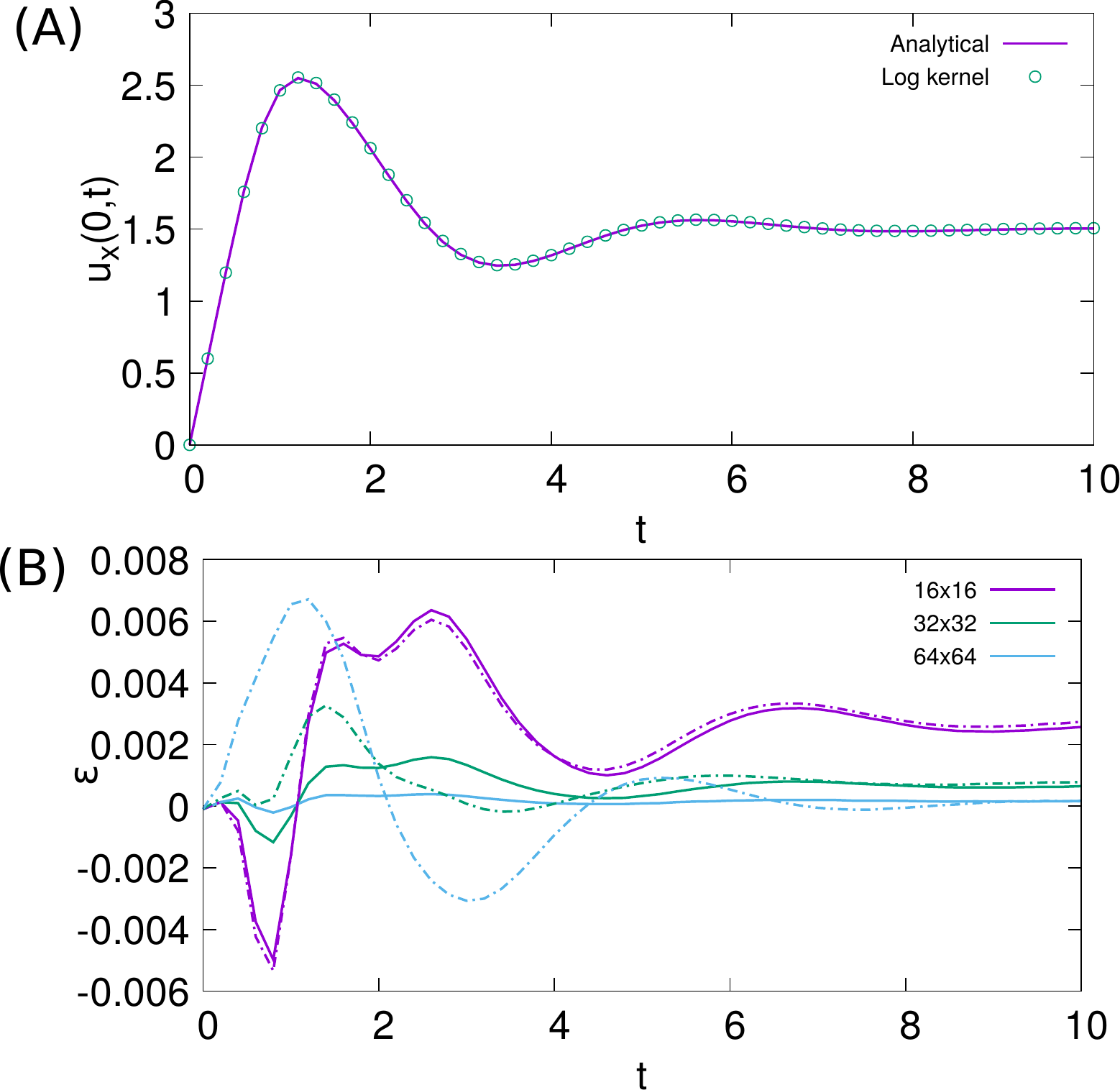}\\
	\caption{Subplot (A): Time evolution of axial velocity $u(0,t)$. Comparison between the analytical solution given by Eq. (\ref{pois_anal}) and the numerical solution with a uniform grid of $16 \times 16$ and a timestep $\Delta t = 10^{-3}$. Subplot (B): Error evolution for three different grids. The continuous line is obtained with a timestep $\Delta t = 10^{-5}$ while the dash-point line is obtained with $\Delta t = 10^{-3}$.  }\label{fig3}
\end{figure}

Using the scaling described above, i.e. $H$, $\lambda$, and $\bar{u}^*_\infty$ for lengths, times and velocities, respectively, the problem is characterized only by the dimensionless magnitudes $E$ and $\beta$ being the dimensionless drop of pressure given by,
\[
\frac{\Delta p }{\Delta x} = \frac{\lambda}{\rho \bar{u}^*_\infty} \frac{\Delta p^*}{\Delta x^*} = -3 E\, .
\]
Therefore, the numerical simulation domain is a square box of dimensionless size $1\times 1$. At the top boundary we set a no-slip condition, while for the bottom symmetry conditions apply. For the left and right boundaries periodic boundary conditions are used for all variables except for the pressure, which is set to $3E$ at the left side and to 0 at the right side.

We have simulated, using the log kernel approach, the case corresponding to $E=1$ and $\beta = 1/9$ with three uniform grids with a dimensionless cell size of $h = 0.0625$  ($16 \times 16$ grid), 0.03125 ($32 \times 32$) and  0.015625  ($64 \times 64$) using a constant time step of value $\Delta t = 0.001$. To use a larger time step compromises the convergence. Subplot A of figure \ref{fig3} illustrates a comparison between the analytical solution given by Eq. (\ref{pois_anal}) with the numerical results obtained with the coarsest grid. As it can be observed, the agreement is very good and comparable to similar schemes \cite{Habla2014}, although the time step in the aforementioned work seems to be smaller. Subplot B illustrates the difference between the theory and the numerical simulation as time proceeds, $\varepsilon(t) = u(0,t)|_{theo} -u(0,t)|_{sim}$ for the three grids reported and two different timesteps; $\Delta t = 10^{-5}$ (continuous line) and $\Delta t = 10^{-3}$ (dash-point line). The refinement of the grid becomes apparent for $\Delta t = 10^{-3}$ when the stationary solution is reached. For $t=15$ the error with the coarsest grid is $2.82 \times 10 ^{-3}$ dropping to  $8.98 \,\times\, 10^{-4}$,  and  to $2.78\,\times\, 10^{-4}$, after each doubling of the spatial resolution. As expected, the error drops with the grid size accordingly to a second-order relation.

\begin{figure}
	\centering
	\includegraphics[width=0.8\linewidth]{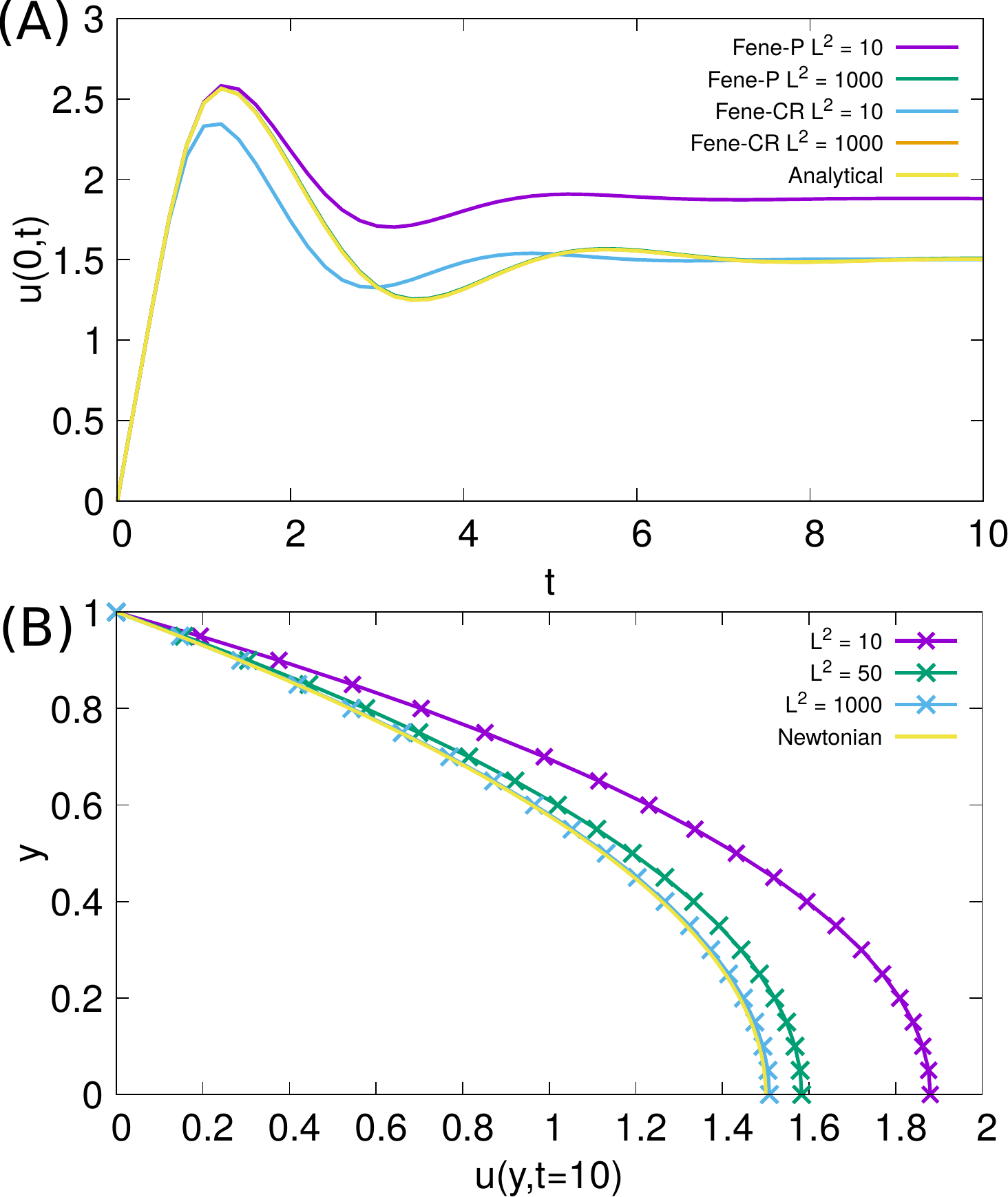}\\
	\caption{Subplot A: Time evolution of the velocity on the axis for FENE-P and FENE-CR. ($L^2$ = 10 and 1000). The Oldroyd-B analytical solution given by Eq.(\ref{pois_anal}) is also shown. Subplot B: Stationary velocity profile for a FENE-P fluid for $L^2=$ 10, 50 and 1000. The stationary solution is supposed to be reached at $t=10$.}\label{fene1}
\end{figure}

The dependence on the viscoelastic model can be observed in figure \ref{fene1}. Subplot A illustrates the temporal evolution of the axial velocity on the axis $u(0,t)$ {\color{black} For two values of the parameter $L^2$, $L^2 = 10$ and $L^2 = 1000$. For each value of the parameter $L$ calculations has been carried out with the FENE-P and  the FENE-CR model}. The analytical solution for Oldroyd-B, Eq. (\ref{pois_anal}), is also shown. Subplot B illustrates the almost stationary velocity profiles for FENE-P with $L^2$ = 10, 50 and 1000. As expected the stationary profiles of FENE-CR coincide with the Newtonian parabolic profile. In contrast, the same pressure gradient creates in a FENE-P fluid a larger average velocity (or flowrate) \cite{Oliveira2002}. For $L^2 \rightarrow \infty$ both FENE-CR and FENE-P coincide with Oldroyd-B. However, the plots in Fig. \ref{fene1} show that, in practice, a value $L^2 = 1000$ suffices.

\begin{figure}
	\centering
	\includegraphics[width=\linewidth]{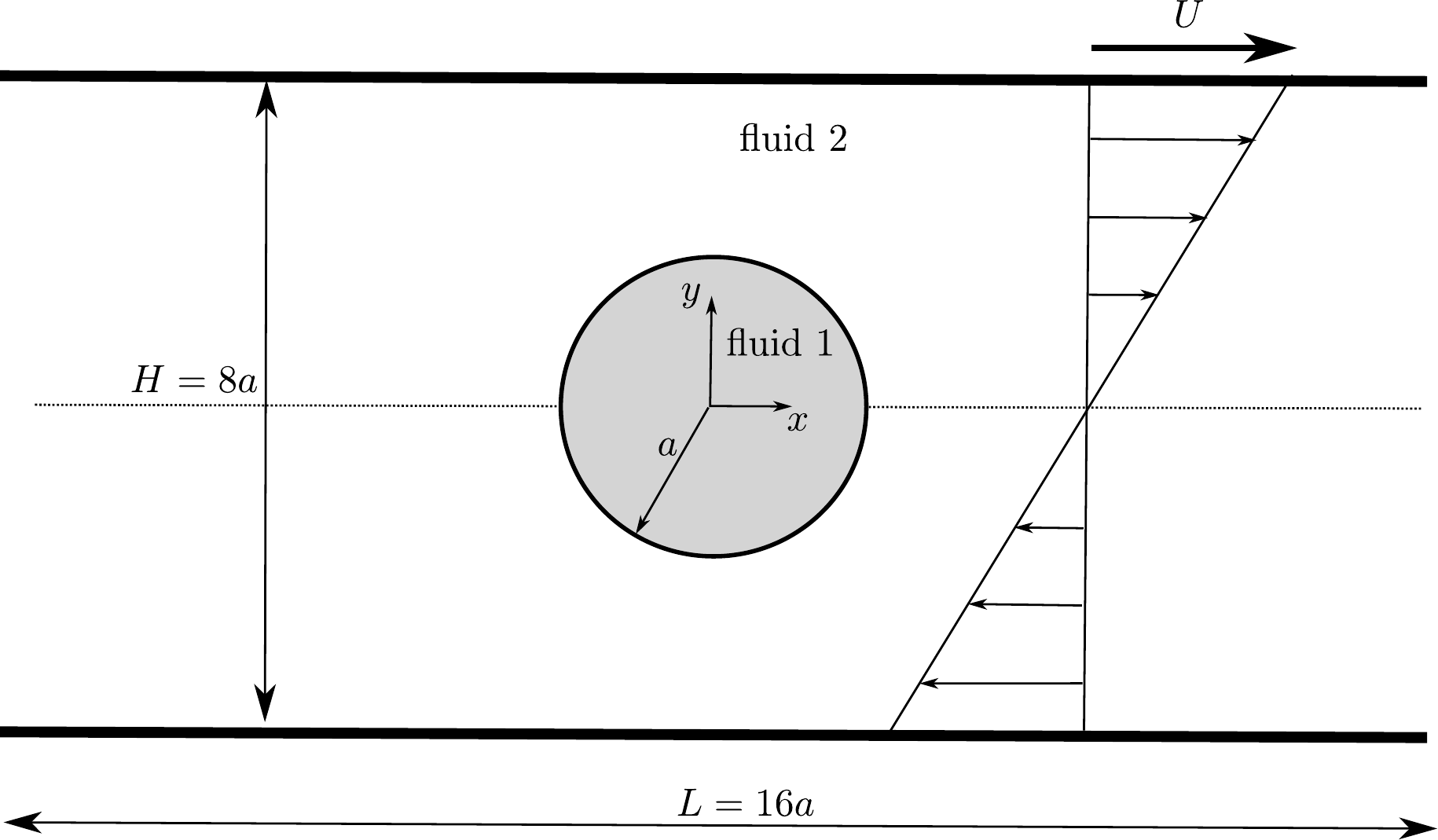}\\
	\caption{Sketch of the benchmark problem of a viscoelastic 2D droplet immersed in a Newtonian fluid undergoing a Couette flow.}\label{figcouette}
\end{figure}
\subsection{2D viscoelastic Oldroyd-B droplet immersed in a Couette  flow}
With this test we wish to validate our scheme using the log kernel methodology when an interface, separating a Newtonian fluid from a viscoelastic one, exists in the presence of surface tension. A sketch of the problem is shown in Fig. \ref{figcouette}. A drop of radius $a$ of the viscoelastic fluid (whose properties we label with the subscript 1) is surrounded by a Newtonian fluid of density and viscosity, $\rho_2$ and $\mu_2$, respectively.  The interfacial surface tension is $\sigma$. Both fluids are trapped, as shown, in a planar gap of a width  equal to eight times the droplet radius $H= 8 a$, and a length approximately sixteen times the droplet radius $L=16 a$. Suddenly a Couette flow is imposed to both fluids,
\[
u^*_x(x^*,y^*;t^* = 0) = \dot{\gamma} \, y^* \quad \mbox{being} \quad \dot{\gamma} = 2U/H \, .
\]
where the superscript $^*$ denotes dimensional variables. {\color{black} The rest of variables are zero initially.} Usually equations are made dimensionless with the outer
density, $\rho_2$, the droplet radius, $a$, and the shear rate,
$\dot{\gamma}$. With this nondimensionalization, the governing parameters of the problem are: the Weber number $We$, the outer Reynolds number, $Re$, the ratio of viscosities and densities, $\mu_r$ and $\rho_r$, the Deborah number $De$, and the ratio of solvent to the total viscosity $\beta$ given by the following expressions,.
\[
We = \frac{\rho_2 a^3 \dot{\gamma}^2}{\sigma}, \:
Re = \frac{\rho_2 a^2 \dot{\gamma}}{\mu_2}, \:
\mu_r =\frac{\mu_1}{\mu_2} , \:
\rho_r = \frac{\rho_1}{\rho_2}, \: De = \dot{\gamma}\, \lambda \quad \mbox{and} \quad \beta = \frac{\mu_s}{\mu_1}\, .
\]
Note that the polymer viscosity is, $\mu_p=\mu_1-\mu_s$ and $\lambda$ is the relaxation parameter. Also, the dimensionless time is, $t = \dot{\gamma}\, t^*$.

This problem was first investigated by \cite{Chinyoka2005} and used as a test problem by many others, see for example \cite{Figueiredo2016, Khismatullin2006, Li2012, Zainali2013}. In Chinyoka et al. \cite{Chinyoka2005} diverse configurations are explored related to the viscoelastic/Newtonian nature of the outer/inner fluid. Since our objective here is to check how our implementation of the log conformation kernel performs in the presence of a fluid interface, we focus on the configuration with an outer Newtonian fluid surrounded by a viscoelastic Oldroyd-B drop. Other configurations have not been considered. In this test we will compare it with the recent results of  Figueiredo et al. \cite{Figueiredo2016}. Therefore, the following characteristic values are set; $Re=0.3$, $We=0.18$, $\mu_r=\rho_r=1$, $De = 0.4$ and $\beta=0.5$. An uniform grid with cells of width $h/a= 3.125 \, \times \, 10^{-2}$ has been used, while Figueiredo et al. used two grids which are not uniform, with a minimum size $h/a= 4.6876 \, \times \, 10^{-2}$ (M1) and $h/a= 2.3438 \, \times \, 10^{-2}$ (M2).

\begin{figure}
	\centering
	\includegraphics[width=\linewidth]{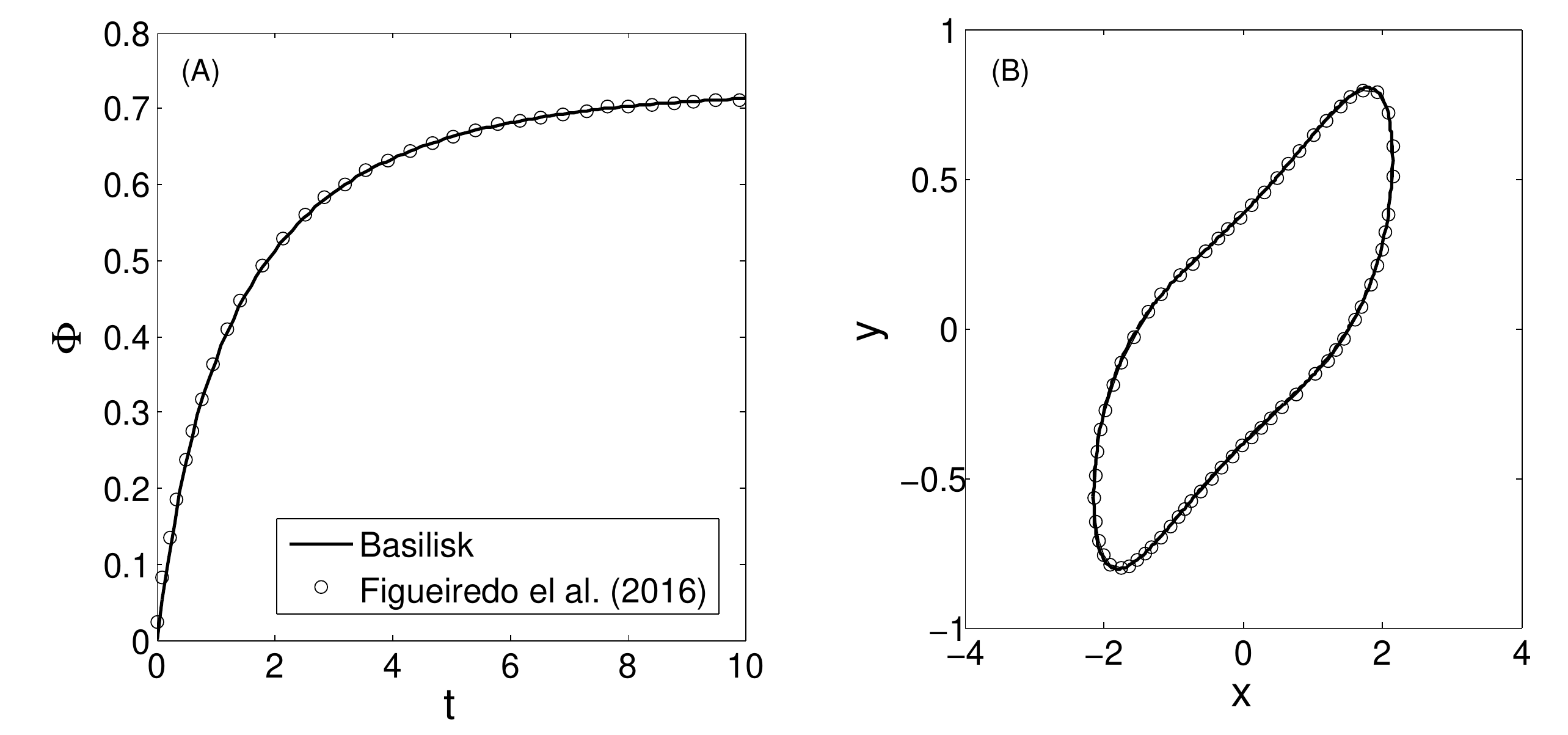}\\
	\caption{Plot (A): Deformation $\Phi$ versus the dimensionless time $t$. Plot (B): deformation of the droplet at time $t=10$. Dimensionless values of the tested case: $Re=0.3$, $We=0.18$, $\mu_r=\rho_r=1$, $De = 0.4$ and $\beta=0.5$. In both plots the results of the proposed scheme are shown with a continuous black line labelled with the name 'Basilisk'. The numerical simulation has been performed with the log kernel methodology. The open circles correspond to the results of \cite{Figueiredo2016}.}\label{figdrop}
\end{figure}

To compare the time evolution of the interface, Chinyoka et al. proposed, as a measuring parameter of the deformation, $\Phi$, the following ratio
\[
\Phi = \frac{R_{max} -R_{min}}{R_{max} + R_{min}}
\]
where $R_{min}$ and $R_{max}$ are, respectively, the minimum and maximum distance between the interface and the droplet center (the origin in our case). {\color{black}This parameter is also known as the Taylor deformation parameter being denoted by $D$}.  Fig. \ref{figdrop}A shows how this parameter evolves in our simulation (black continuous line labelled as 'Basilisk') compared with Figueiredo et al. \cite{Figueiredo2016} (open circles). Also, Fig. \ref{figdrop}B shows the position of the interface for both simulations. As expected, the agreement between both simulations is excellent.

\begin{figure}
	\centering
	\includegraphics[width=\linewidth]{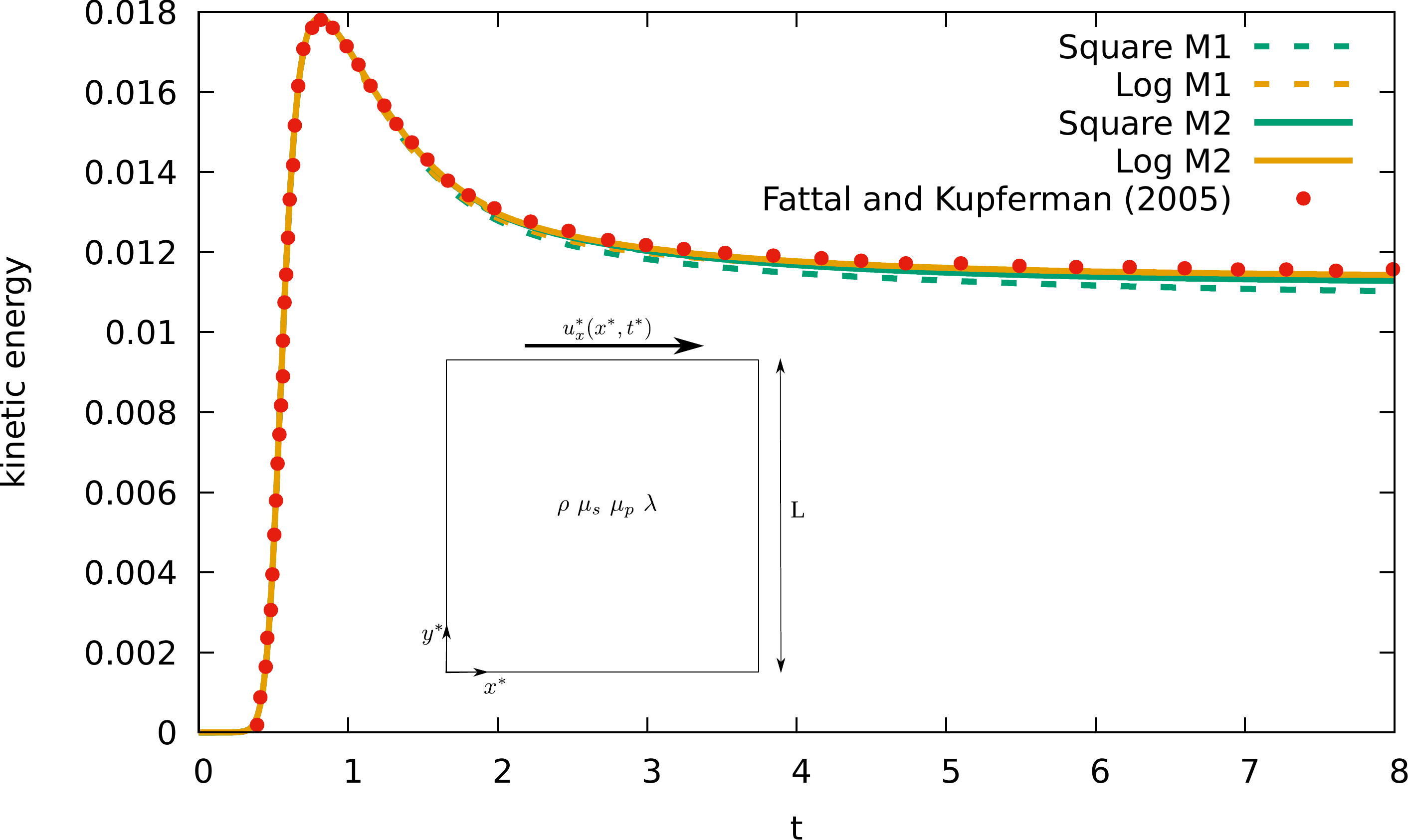}\\
	\caption{Time evolution of the dimensionless kinetic energy for $Wi = 1$, $\beta = 0.5$ and $Re=0.01$. The simulations have been performed with two uniform grids; M1 and M2 (64$\times$64 and 128$\times$128, respectively) and an adapted grid. Numerical results with (i) the log-conformation kernel and (ii) the square root kernel are shown. Results of Fattal \& Kupferman (2005) are also shown (red circles). Insert: Sketch of the lid cavity problem.}\label{lid1}
\end{figure}

\begin{figure}
	\centering
	\includegraphics[width=0.7\linewidth]{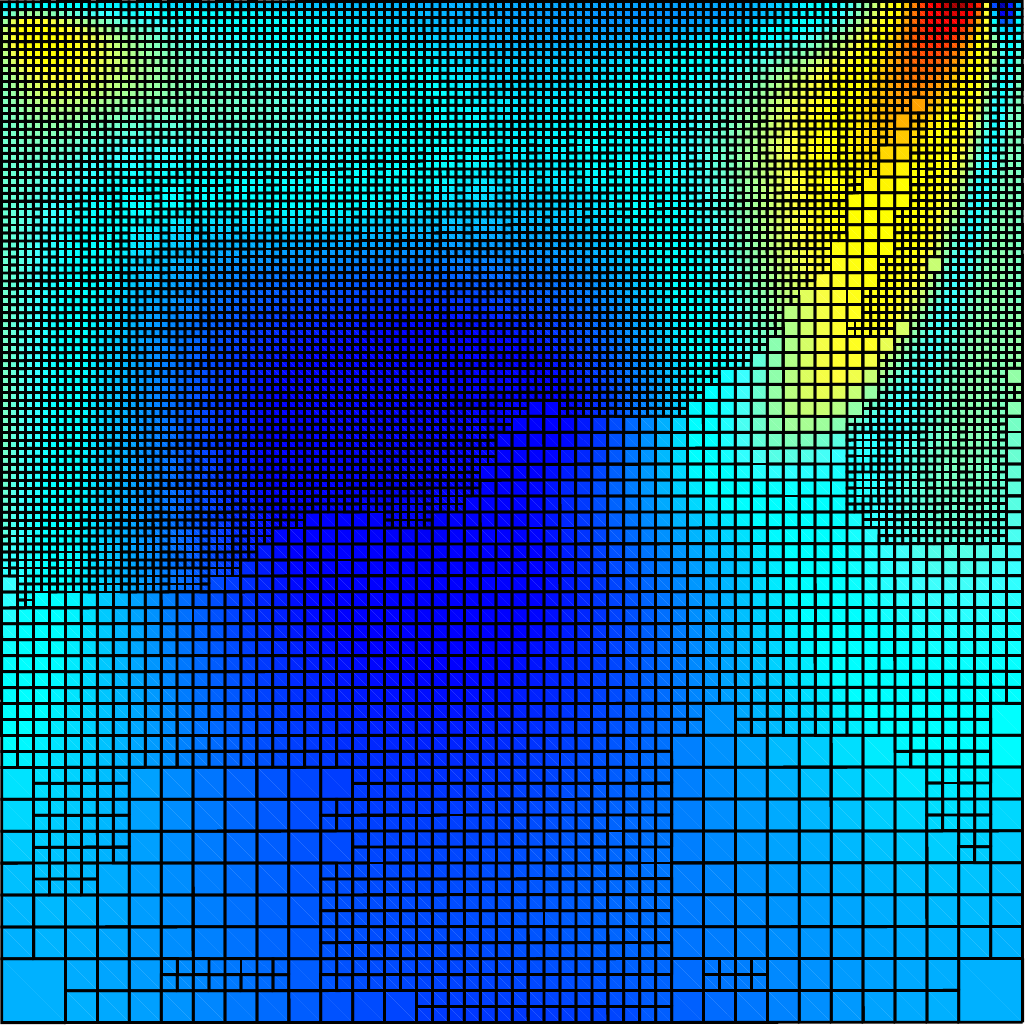}\\
	\caption{Spatial distribution of $\bm{\Psi}_{xy}$ at instant $t=3$. The adapted grid at that instant is also shown.}\label{lid1a}
\end{figure}

\begin{figure}
	\centering
	\includegraphics[width=0.7\linewidth]{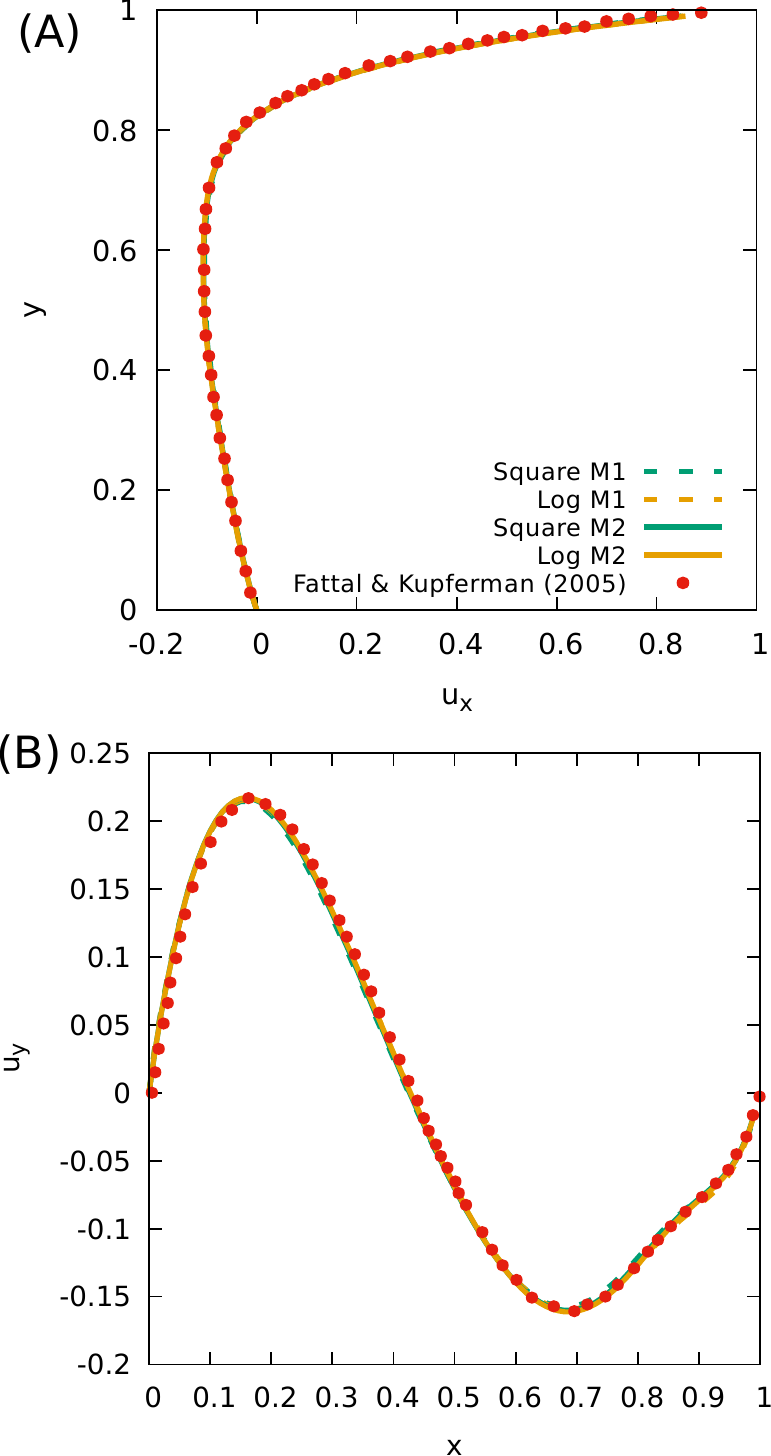}\\
	\caption{Subplot A: Profile of the x-component of the velocity, $u_x$, at position $x=0.5$. Subplot B: Profile of the y-component of the velocity, $u_y$, at position $y=0.75$. The profiles correspond to instant $t=8$ and the fluid parameters are $Wi = 1$, $\beta = 0.5$ and $Re=0.01$. The legend of the curves is the same as that in subplot A.}\label{lid2}
\end{figure}

\subsection{lid cavity flow}
This test deals with the movement of a viscoelastic Oldroyd-B fluid of density $\rho$, relaxation parameter $\lambda$, and solvent and polymeric viscosities, $\mu_s$ and $\mu_p$, respectively. As shown in the insert of Fig \ref{lid1}, the fluid is confined in a square cavity of size $L$, bounded by walls, except on the top side where a time-dependent tangential velocity is imposed. Using as scaling magnitudes the density $\rho$, the largest stationary velocity $U_o$ and the width of the cavity $L$, we form a Weissenberg number, $Wi$, a Reynolds number, $Re$, and a solvent viscosity ratio, $\beta$, given by
\[
Wi = \frac{\lambda L}{U_o}, \quad Re = \frac{\rho U_o L}{\mu_o} \quad \text{and} \quad \beta = \frac{\mu_s}{\mu_o} \, \quad \text{where} \quad \mu_o = \mu_p +\mu_s \, .
\]
The standard problem relies on the following regularized dimensionless parabolic profile for the top lid
\begin{equation*}
u_x (x, t) = 8 \left[ 1 + \tanh \left( 8t - 4 \right)\right]x^2 (1-x)^2
\end{equation*}
where $x=x^*/L$, $t= U_o t^*/L$ and $u_x = u^*_x/U_o$ are the corresponding dimensionless variables. The remaining cavity walls are stationary and the no-slip boundary condition is imposed on the four walls. We assume the Stokes limit for the momentum equation. In the simulation, we have set $Re = 0.01$, $\beta$ = 0.5 and $Wi = 1$. This test case has become a classical benchmark problem in computational rheology since the HPWN manifests itself with these values of the dimensionless parameters. In Table 1 of \cite{Sousa2016} are gathered previous numerical studies concerned with a lid-driven cavity flow of constant viscosity viscoelastic fluids. We have solved this test case with a uniform grid of 64 $\times$ 64 (grid M1) and with a grid of $128\times 128$ (grid M2) equivalent to a \emph{level} $\ell=6$ and 7, respectively. The maximum timestep for the M1 grid is $\Delta t = 5\,\times \, 10^{-5}$, while for M2 we had to set $\Delta t =  10^{-5}$. Numerical simulations with an adapted grid have  also been carried out. The adaptation is applied every 50 timesteps by controlling the error on the components of the dimensionless velocity. The threshold value for both components is $5 \, \times \, 10^{-4}$  with the maximum and minimum levels of refinement/coarsening, $\ell=7$ and 5, respectively. Figure \ref{lid1} shows the time evolution of the total dimensionless kinetic energy in the cavity,
\begin{equation*}
\frac{1}{2} \int_0^1 \int_0^1 (u^2_x + u^2_y) \, dx \, dy
\end{equation*}
The simulation obtained with the log-conformation kernel, the square root kernel, and the results of Fattal \& Kupferman (2005) are shown. The dashed line indicates that the results were obtained with grid M1. A continuous line corresponds to grid M2. The results of Fattal \& Kupferman were obtained with a grid of 256$\times$256. We also plot in figure \ref{lid2} velocity profiles at instant $t=8$. In subplot A we show the profile $u_x$ at the position $x=0.5$, while in subplot B is shown the profile $u_y$ at the height $x=0.75$. As in Fig. \ref{lid1}B we show either the simulations performed with the log conformation kernel, the square root kernel and the results of Fattal \& Kupferman.

The agreement in the velocity profiles between the different methodologies, and the previous work shown in Fig. \ref{lid2}, is excellent, although, this is a common result in other schemes. The agreement in the kinetic energy is also very good. In particular, the agreement with the position ($t \sim 0.8$) and intensity (equal to approx. $0.0178$) of the peak of kinetic energy is excellent. However, the square root kernel has a stationary value ($\sim 0.011022$) below  that obtained with the log kernel ($\sim 0.011337$) for the M1 grid. When the grid is doubled, i.e. grid M2, the result for the square root kernel increases to $\sim 0.011282$ and the log kernel to $\sim 0.011429$, closer to the value extracted from \cite{Fattal2005} ($\sim 0.011572$). Adaptation allows  the grid to be refined where needed. In the lid cavity problem, as can be observed in Fig. \ref{lid1a}, refinement is located close to the moving wall. Since the velocity is almost established at instant $t=3$ the grid distribution shown in Fig. \ref{lid1a} changes little in later instants.  It seems that the log kernel gives a slightly more accurate result than the square root kernel. Interestingly, a similar trend can be observed in Figure 4.b of \cite{PalharesJunior2016}. It is worth mentioning that for the calculation of the lid cavity problems, Figueiredo et al. \cite{Figueiredo2016} report timesteps of $10^{-4}$, about an order of magnitude larger than ours.

\end{document}